\documentclass[5p,times]{elsarticle}

\usepackage{todonotes}

\usepackage{ecrc}

\volume{00}

\firstpage{1}

\journalname{Procedia Computer Science}

\runauth{}

\jid{procs}

\jnltitlelogo{Procedia Computer Science}

\CopyrightLine{2011}{Published by Elsevier Ltd.}

\usepackage{amssymb}

\usepackage[figuresright]{rotating}

\usepackage{balance}
\usepackage{graphicx}
\usepackage{multirow}
\usepackage{multicol}
\usepackage{float}
\usepackage{xcolor}
\usepackage{amssymb}
\usepackage{amsmath}
\usepackage{minibox}
\usepackage{tabularx}
\usepackage{caption}
\usepackage{subcaption}
\usepackage{pgf}
\usepackage{enumitem}
\usepackage{array}
\usepackage{csquotes}
\usepackage[hyphens]{url}

\usepackage{listings}
\newcommand{\cylon}{\textit{Cylon}}

\begin{document}

\begin{frontmatter}



\dochead{}

\title{In-depth Analysis On Parallel Processing Patterns for High-Performance Dataframes}


\author[iu]{Niranda Perera}
\ead{niranda@niranda.dev}

\author[uva,uvab]{Arup Kumar Sarker}
\ead{djy8hg@virginia.edu}

\author[uva]{Mills Staylor}
\ead{qad5gv@virginia.edu}

\author[uvab]{Gregor von Laszewski}
\ead{laszewski@gmail.com}

\author[uva]{Kaiying Shan}
\ead{shankaiying@gmail.com}

\author[iu]{Supun Kamburugamuve}
\ead{supun@apache.org}

\author[iu]{Chathura Widanage}
\ead{chathurawidanage@gmail.com}

\author[iu]{Vibhatha Abeykoon}
\ead{vibhatha@gmail.com}

\author[iu]{Thejaka Amila Kanewela}
\ead{thejaka.amila@gmail.com}

\author[uva,uvab]{Geoffrey Fox}
\ead{vxj6mb@virginia.edu}

\address[iu]{Indiana University Alumni, Bloomington, IN 47405, USA }
\address[uva]{University of Virginia, Charlottesville, VA 22904, USA}
\address[uvab]{Biocomplexity Institute and Initiative, University of Virginia, Charlottesville, VA 22904, USA}

\begin{abstract}


The Data Science domain has expanded monumentally in both research and industry communities during the past decade, predominantly owing to the \textit{Big Data} revolution. Artificial Intelligence (AI) and Machine Learning (ML) are bringing more complexities to data engineering applications, which are now integrated into data processing pipelines to process terabytes of data. Typically, a significant amount of time is spent on data preprocessing in these pipelines, and hence improving its efficiency directly impacts the overall pipeline performance. The community has recently embraced the concept of \textit{Dataframes} as the de-facto data structure for data representation and manipulation. However, the most widely used serial Dataframes today (R, \texttt{pandas}) experience performance limitations while working on even moderately large data sets. We believe that there is plenty of room for improvement by taking a look at this problem from a high-performance computing point of view. In a prior publication, we presented a set of parallel processing patterns for distributed dataframe operators and the reference runtime implementation, \cylon{} \cite{perera2022high}. In this paper, we are expanding on the initial concept by introducing a cost model for evaluating the said patterns. Furthermore, we evaluate the performance of \cylon{} on the ORNL Summit supercomputer.

\end{abstract}


\end{frontmatter}








\section{Introduction}

Artificial Intelligence (AI), Machine Learning (ML), and the \textit{Big Data} revolution have introduced an abundance of complex data engineering applications in the data science domain. These applications are now required to process terabytes of data and are orchestrated as an intricate collection of data engineering pipelines. To achieve this, a significant amount of \textit{developer time} is spent on data exploration, preprocessing, and prototyping. Therefore, improving the efficiency of such activities directly impacts the overall data engineering pipeline performance.

Databases and structured query language (SQL) have been the de-facto tool for data preprocessing applications. However, in the early 2000s, the focus shifted significantly towards \textit{Big Data} toolkits and frameworks. These systems (eg. Hadoop \cite{hadoop:online} and map-reduce \cite{dean2008mapreduce}, Spark \cite{spark:online}, Flink \cite{flink:online}, etc.) enabled more capabilities than traditional relational database management systems (RDBMS), such as functional programming interface, consuming large structured and unstructured data volumes, deploying in the cloud at scale, etc. Coinciding with the big data developments, enterprise and research communities have invested significantly in artificial intelligence and machine learning (AI/ML) systems. Data analytics frameworks complement AI/ML by providing a rich ecosystem for preprocessing data, as these applications require enormous amounts of 
data to train their models properly.

In recent times, the data science community has increasingly moved away from established SQL-based abstractions and adopted Python/R-based approaches, due to their user-friendly programming environment, optimized execution backends, broad community support, etc. \textit{Dataframes} play a pivotal role in this transformation \cite{mckinney2011pandas} by providing a functional interface and interactive development environment for exploratory data analytics. Most dataframe systems available today (e.g. R-dataframe, Pandas) are driven by the open-source community. However, despite this popularity, many dataframe systems encounter performance limitations even on moderately large data sets. 
We believe that dataframe systems have now exhausted the capabilities of a single computer and this paves the way for distributed and parallel dataframe processing systems.

    \subsection{Background: High-Performance Dataframes from Parallel Processing Patterns}
    \begin{sloppypar}
    In the precursor publication, titled "High-Performance Dataframes from Parallel Processing Patterns" \cite{perera2022high}, we presented a framework that lays the foundation for building high-performance distributed-memory parallel dataframe systems based on parallel processing patterns. There, we analyzed the semantics of common dataframe operators to establish a set of generic distributed operator patterns. We also discussed several significant engineering challenges related to developing a scalable and high-performance distributed dataframe (DDF) system. The main goal of this framework is to simplify the DDF development process substantially by promoting existing serial/ local operators into distributed operators following the said patterns. They primarily focus on a distributed memory and Bulk Synchronous Parallel (BSP) \cite{valiant1990bridging,fox1989solving} execution environment. This combination has been widely employed by the high-performance computing (HPC) community for exascale computing applications with admirable success. Based on this framework, we developed \cylon{}, an open-source high-performance distributed dataframe system \cite{cylon}. 
    \end{sloppypar}

\vspace{1em}
In this paper, we present an in-depth analysis of the aforementioned parallel processing patterns based on a cost model. We encapsulate the parallel processing patterns concept into "\cylon{} Distributed Operator Model" and present "\cylon{} Communication Model" which allows plugging-in multiple communication runtimes into \cylon{} distributed execution. These two aspects constitute the "\cylon{} Distributed Memory Execution Model", which we will discuss in detail in the following sections. Furthermore, we will introduce a cost model to evaluate the performance of distributed memory execution. In addition, we demonstrate the scalability of \cylon{} on leadership-class supercomputing environments, which affirms the significance of the underlying framework. We have also conducted a scalability analysis between \cylon{} and related state-of-the-art data processing systems. This analysis demonstrates the applicability of the design across the board, on both distributed computing and supercomputing infrastructure. In the following sections, we use \cylon{} to refer to its underlying high-performance DDF framework interchangeably.  

\section{\cylon{} Distributed-Memory Execution Model} \label{sec:mem_model}

\cylon{} is based on the \textit{distributed memory parallel} model, which isolates memory for each parallel process. These processes can manage their memory individually while communicating with others using message passing. This isolation makes distributed operator implementation easier to reason about. While it leaves room for improvement, especially using multi-threading execution, the results show that \cylon{} dataframes show superior scalability over the state-of-the-art systems. 
In addition, it is based on BSP execution in the distributed memory environment. Gao et al. \cite{gao2021scaling} recently published a similar concept for scaling joins over thousands of Nvidia Graphical Processor Units (GPU). \cylon{} experiments demonstrate that this approach can be generalized to all operators and achieves commendable performance. 

Conceptually, we can divide \cylon{} distributed execution model into two distinct sub-models, \textbf{1. Communication Model}, and \textbf{2. Distributed Operator Model}. We will discuss the former in Section \ref{sec:comm_model} and the latter in Section \ref{sec:op_model}. 

   \subsection{Distributed Memory Parallel Dataframe Definition}

    \begin{figure}[htpb]
    \begin{center}
    \includegraphics[width=\linewidth]{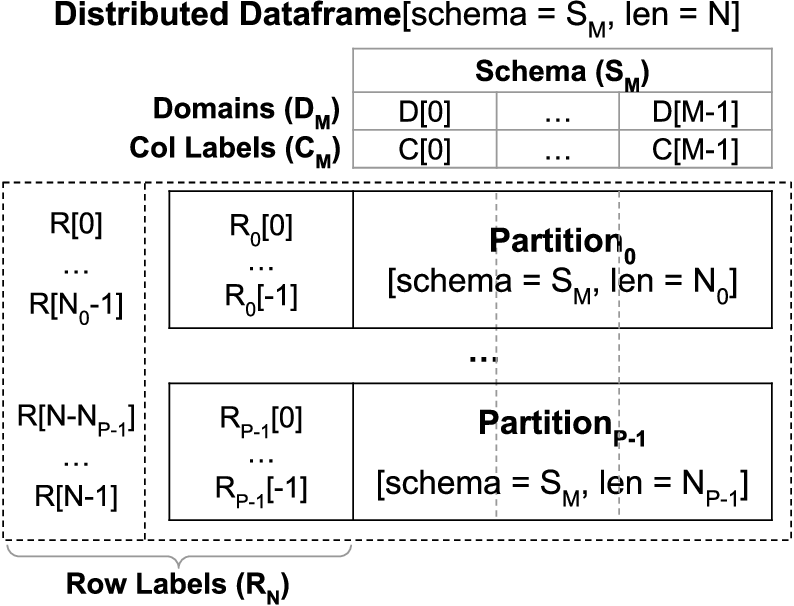}
    \end{center}
    \caption{Distributed Memory Dataframe Abstraction}
    \label{fig:dist-df}
    \end{figure}    

    The primary insight behind \cylon{} is to present a dataframe framework that promotes an already available \textit{serial (local) operator} into a distributed memory parallel execution environment \cite{widanage2020high}. For this purpose, we formally defined a Distributed Memory Parallel Dataframe based on row-based partitioning in our previous publication \cite{perera2022high}. This concept is depicted in Figure \ref{fig:dist-df}. The dotted lines represent the \textit{virtual} collection of \textit{Partitions} in the distributed memory parallel environment. Users would not see a separate distributed API object but instead, continue to write their program as they would work on a single partition. The execution environment determines if the operator needs to be performed locally or in a distributed fashion based on the operator's semantics.

    For example, Figure \ref{fig:pandas-ex} shows a Pandas script that reads data from two directories, joins them, sorts the result, and takes the top 10 rows. A corresponding \cylon{} script for distributed-memory Dataframes is shown in Figure \ref{fig:cylon-ex}. 

    \begin{figure*}[h]
    \centering
    \begin{subfigure}{.4\textwidth}
      \centering
              \begin{lstlisting}
    df1 = read_csv('dir/path/0') #read
    df2 = read_csv('dir/path/1')
    
    df_j = df1.merge(df2, ...) #join
    df_s = df_j.sort_values(...) #sort
    df_s.iloc[:10] # head(10)
        \end{lstlisting}
      \caption{Pandas}
      \label{fig:pandas-ex}
    \end{subfigure}%
    \begin{subfigure}{.6\textwidth}
      \centering
        \begin{lstlisting}
    df1 = read_csv_dist('dir/path/0', env=env) #dist read
    df2 = read_csv_dist('dir/path/1', env=env) 
    
    df_j = df1.merge(df2, ..., env=env) #dist join
    df_s = df_j.sort_values(..., env=env) #dist sort
    df_s.iloc[:10, env] #dist head(10)
        \end{lstlisting}      
        \caption{Cylon}
      \label{fig:cylon-ex}
    \end{subfigure}
    \caption{Example script}
    \label{fig:example}
    \end{figure*}

    \subsection{Apache Arrow Columnar Memory Layout}
    \cylon{} uses Apache Arrow Columnar format as the physical data representation. This is an integral component of the \cylon{} memory model. It provides several benefits, such as data adjacency for sequential access (scans), $O(1)$ (constant-time) random access, SIMD vectorization-friendly data structure, true zero-copy access in shared memory, etc. It also allows serialization-free data access from many language runtimes. Due to these benefits, many libraries including Pandas, PySpark \cite{spark:online}, CuDF \cite{cudf}, and Ray \cite{moritz2018ray}, are now using the Apache Arrow format.    
    
    
    
    
    


\section{\cylon{} Communication Model} \label{sec:comm_model}

In many dataframe applications, communication operations take up significant time creating critical bottlenecks. This is evident from our experiments (Section \ref{sec:experiments}), where we evaluate communication and computation time breakdown applied to several dataframe operator patterns. Moreover, most frameworks (eg. Spark, Dask, Ray), provide special guidelines to reduce communication overheads (eg. shuffle routine) \cite{daskshuffle:online,rayshuffle:online}. Therefore, careful attention has been given while developing the communication model for \cylon{}. 

BSP execution allows the program to continue independently until the next communication boundary is reached. Message passing libraries such as MPI (OpenMPI, MPICH, IBM Spectrum, etc), Gloo, and UCX \cite{shamis2015ucx} provide communication routines for memory buffers, which by extension support homogeneously typed arrays. The most primitive routines are point-to-point (P2P) message passing, i.e., tag-based \textit{async send} and \textit{async receive}. Complex patterns (generally termed \textit{collectives}) can be derived on top of these two primitive routines (eg. MPI-Collectives, UCX-UCC). 

\begin{table*}[htpb]
\centering
\begin{tabular}{c|ccc|}
\cline{2-4}
\multicolumn{1}{l|}{}                    & \multicolumn{3}{c|}{\textbf{Data Structure}}                                                \\ \hline
\multicolumn{1}{|c|}{\textbf{Operation}} & \multicolumn{1}{c|}{\textbf{Table}} & \multicolumn{1}{c|}{\textbf{Array}} & \textbf{Scalar} \\ \hline
\multicolumn{1}{|c|}{Send/ Recv} & \multicolumn{1}{c|}{Common}         & \multicolumn{1}{c|}{Common}           & Common             \\ \hline
\multicolumn{1}{|c|}{Shuffle (AllToAll)} & \multicolumn{1}{c|}{Common}         & \multicolumn{1}{c|}{Rare}           & N/A             \\ \hline
\multicolumn{1}{|c|}{Scatter}            & \multicolumn{1}{c|}{Common}         & \multicolumn{1}{c|}{Rare}           & N/A             \\ \hline
\multicolumn{1}{|c|}{Gather/AllGather}  & \multicolumn{1}{c|}{Common}         & \multicolumn{1}{c|}{Common}         & Common          \\ \hline
\multicolumn{1}{|c|}{Broadcast}          & \multicolumn{1}{c|}{Common}         & \multicolumn{1}{c|}{Common}         & Common          \\ \hline
\multicolumn{1}{|c|}{Reduce/AllReduce}  & \multicolumn{1}{c|}{N/A}            & \multicolumn{1}{c|}{Common}         & Common          \\ \hline
\multicolumn{1}{|c|}{Barrier}            & \multicolumn{3}{c|}{Common (independent of the data structure)} \\ \hline
\end{tabular}
\caption{Communication semantics in Dataframe Operators}
\label{tab:com-semantics}
\end{table*}

Unlike multi-dimensional arrays, heterogeneous data types in dataframes make communication routines more involved. The Arrow columnar data format represents a column by a tuple of buffers (\texttt{boolean} validity bitmap, \texttt{integer} offsets, \& \texttt{byte} data). A dataframe incorporates a collection of such columns. Therefore, a communication routine would have to be called on each of these buffers. \cylon{} communication model outlines a set of communication collectives required to implement distributed memory parallel dataframes by inspecting the semantics of core dataframe operators. These are listed in Table \ref{tab:com-semantics} together with their frequency of usage for each data structure. 

The key features of the \cylon{} communication model are, 

\begin{enumerate}
    \item \textbf{Modular architecture}: Allows plugging-in multiple communication libraries.
    \item \textbf{Extensibility}: The communication model has been easily extended into Nvidia CUDA GPU hardware, in \textit{GCylon} project. 
\end{enumerate}

    \begin{figure}[htpb]
    \begin{center}
    \includegraphics[width=\linewidth]{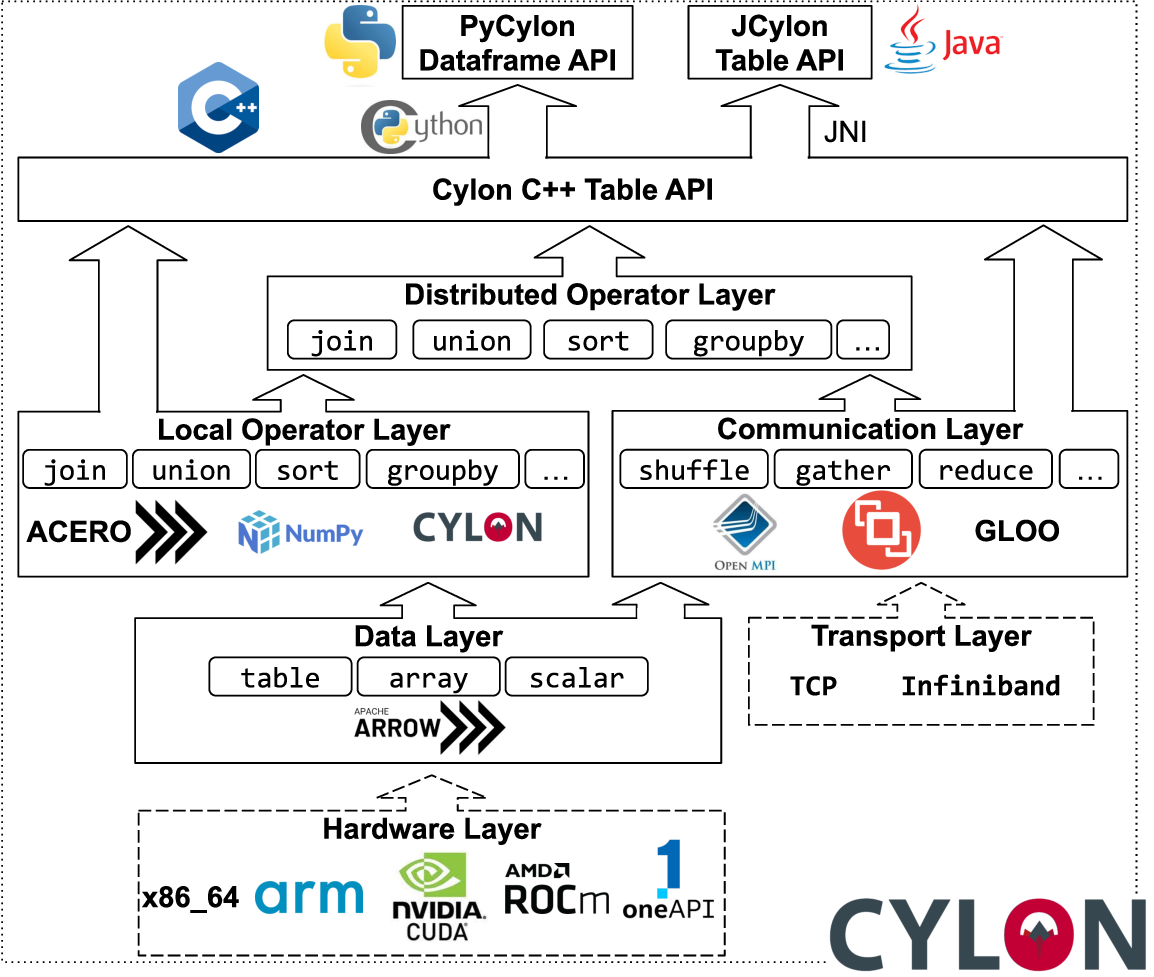}
    \end{center}
    \caption{\cylon{} Architecture}
    \label{fig:cylon_arch}
    \end{figure}

    Figure \ref{fig:cylon_arch} depicts the overall \cylon{} architecture. 

    \subsection{Communicator}
    
\begin{figure}[htpb]
    \begin{center}
    \includegraphics[width=\linewidth]{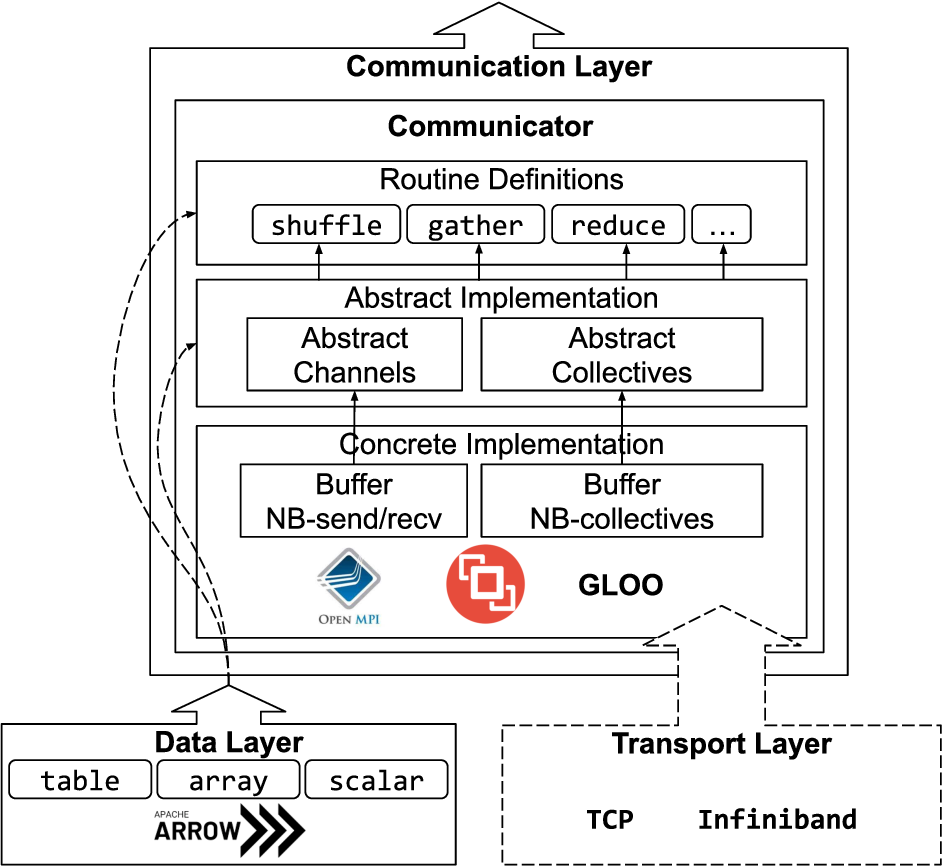}
    \end{center}
    \caption{\cylon{} Communicator Model}
    \label{fig:comm_model}
    \end{figure}

    The \textit{communicator} interface manages \cylon{} communication routines (Figure \ref{fig:comm_model}). At the very top, the user API defines routines based on the data layer data structures, as described in Table \ref{tab:com-semantics}. These are blocking routines for the user (e.g., \texttt{shuffle\_table} will wait until completion).

    The communicator implements these routines using two abstract constructs, (1). channels (for point-to-point/ send-receive communications) and (2). collective communications. The former works only on byte buffers, and the collectives can also be implemented using these channels. In fact, \texttt{table\_shuffle} is implemented using channels due to a mismatch in traditional \texttt{MPI\_Alltoall}. The abstract collective communications implement collective routines for composite data structures (tables, arrays, and scalars), using collectives on buffers. This abstract implementation allows \cylon{} to easily plug in multiple communication libraries that support BSP semantics, such as OpenMPI \cite{OpenMPI:online}, UCX \cite{shamis2015ucx}, and Gloo \cite{gloo:online}.
    
    \subsection{Abstract Channels}
    Channels are designed to be used for composite buffer communications in a non-blocking manner. During the initialization, it registers two callbacks which inform the caller that (1). the sending has been completed, (2). the data is received for a particular buffer. It then accepts \textit{requests} that contain the buffer address and metadata (such as buffer size, buffer index, etc.) to be sent. The caller then has to progress through sends and receives. First, the channel exchanges buffer metadata, which is used to allocate memory for receiving buffers. Later on, it starts exchanging data. Both these progressions use non-blocking send/receive routines. Once each receiving buffer completes, it will be passed on to the caller using the receive-callback. 
    
    Channels give much flexibility to the caller to implement composite communication routines. However, there are disadvantages to this as well. Most importantly, each buffer collective routine must be implemented from scratch using channels. As listed in Table \ref{tab:comm-comp}, we need to implement multiple communication algorithms to get the best performance for collectives. Managing such a custom communication library code base could be a cumbersome exercise. Currently, \texttt{shuffle} routine is implemented using the channels. 

    \subsection{Abstract Collectives}

    \begin{sloppypar}
    Abstract collectives are higher-level communication abstraction that implements table, array, or scalar collectives using non-blocking buffer collective routines. For example, an \texttt{allgather table} can be implemented as a collection of non-blocking \texttt{allgather} routines. To do this, we create a metadata structure with the buffer pointers, sizes, data types, etc. of the input table and call corresponding communication routines on each buffer. In the end, we recreate the resultant table based on the output buffers.         
    \end{sloppypar}
    
    
    \subsection{Supported Communication Libraries} 
    
    Currently, \cylon{} communicator supports the following communication libraries that support BSP message-passing semantics. 
        
        \subsubsection{OpenMPI} 
        OpenMPI is a widely used open-source implementation of the MPI specification. 
        It consists of two main components, (1). process management and (2). communication library. Currently, Process Management Interface Exascale (PMIx) standard \cite{PMIx:online} is used for the former, while various communication algorithms have been implemented (Table \ref{tab:comm-comp}) as a part of the latter. It is a comprehensive communication library with a rich collection of communication routines for many distributed computing and HPC applications. \cylon{} communication model was also heavily influenced by OpenMPI.  
        

        \subsubsection{Gloo}
        \textit{Gloo} collective communications library is managed by Meta Inc. incubator \cite{gloo:online} predominantly aimed at machine learning applications. PyTorch uses it for distributed all-reduce operations. It currently supports TCP, UV, and ibverbs transports. Gloo communication runtime can be initialized using an MPI Communicator or an NFS/Redis key-value store (P2P message passing is not affected). Gloo lacks a comprehensive algorithm implementation as an incubator project, yet our experiments confirmed that it scales admirably. We have extended the Gloo project to suit \cylon{} communication interface.
        
        \subsubsection{UCX/UCC}
        Unified Communication X (UCX) is a collection of libraries and interfaces that provides an efficient and convenient way to construct widely used HPC protocols on high-speed networks, including MPI tag matching, Remote Memory Access (RMA) operations, etc. Unlike MPI runtimes, UCX communication workers are not bound to a process bootstrapping mechanism. As such, it is being used by many frameworks, including Apache Spark and RAPIDS (Dask-CuDF). It provides primitive P2P communication operations. Unified Collective Communications (UCC) is a collective communication operation API built on UCX, which is still being developed. Similar to MPI, UCC implements multiple communication algorithms for collective communications. Based on our experiments, UCX+UCC performance is on par with or better than OpenMPI. 
    
\section{\cylon{} Distributed Operator Model}\label{sec:op_model}

\begin{figure}[htpb]
    \centering
    \begin{tabular}{c}
    \includegraphics[width=\linewidth]{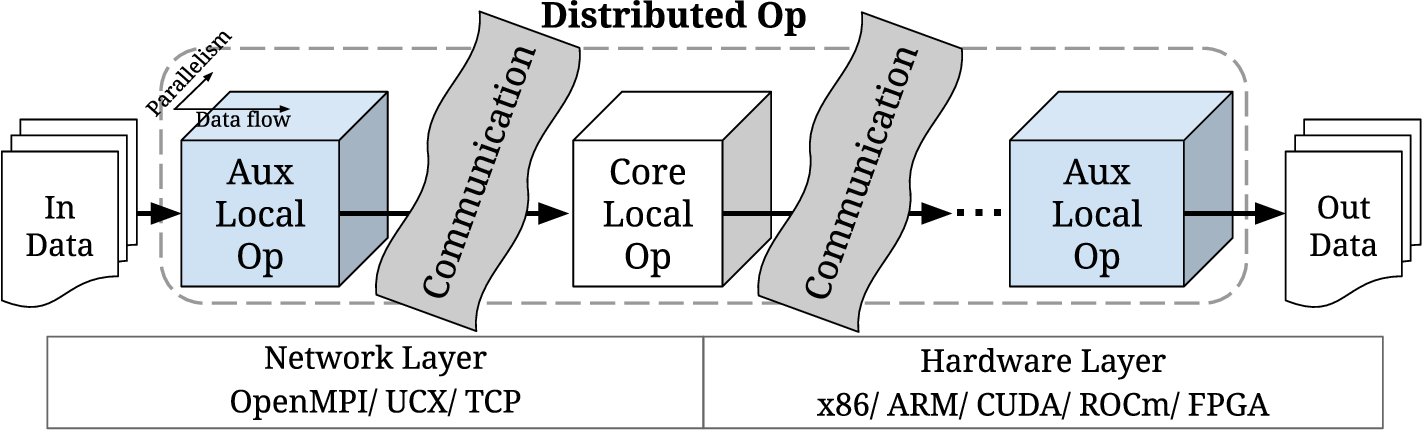}\\
    \includegraphics[width=\linewidth]{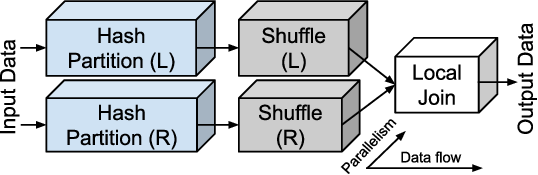}
    \end{tabular} 
    \captionof{figure}{Distributed DDF Sub-operator Composition \cite{perera2022high} (Bottom: Join Operator Example)}
    \label{fig:dist-join}
\end{figure}

\begin{figure}[htpb]
    \begin{center}
    \includegraphics[width=\linewidth]{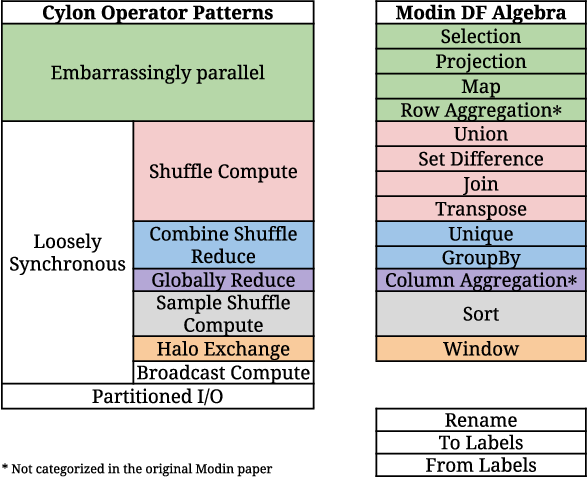}
    \end{center}
    \captionof{figure}{\cylon{} Operator Patterns \& Modin DF Algebra}
    \label{fig:cylon_modin}
\end{figure}

\cylon{} distributed operator model provides the basis for elevating a local dataframe operator to a distributed memory parallel dataframe operator. This was the primary idea behind our precursor publication \cite{perera2022high}. It comprises two key observations,

\begin{enumerate}
    \item A distributed operator consists of three major sub-operators: 
        \begin{enumerate}
            \item Core local operator
            \item Auxiliary local operators
            \item Communication operators
        \end{enumerate}
    For example, the bottom image in Figure \ref{fig:dist-join} shows how the distributed \texttt{join} is composed of these sub-operators.
    \item By examining the composition of these sub-operators, they can be categorized into several parallel execution patterns, as depicted in Figure \ref{fig:cylon_modin}. Therefore, rather than analyzing/ optimizing each operator, we can focus on these parallel patterns. In addition, some operators can be implemented using multiple algorithms that show distinctive parallel patterns (e.g., \texttt{join} can be done by shuffling or by broadcasting). Hence, understanding these patterns is essential to choose the best runtime strategy.  
\end{enumerate}

We believe understanding distributed dataframe operator patterns reduce the burden of parallelizing a massive API, such as Pandas. To address the same problem, Petersohn et al. \cite{petersohn2020towards} introduced a primitive set of dataframe operators that could be used as a \textit{basis} for the rest, termed \textit{Dataframe Algebra}. Our dataframe operator patterns are a complementary concept to dataframe algebra, as shown in Figure \ref{fig:cylon_modin}.

    \subsection{Core Local Operator}
    These refer to single-threaded implementations of primitive operators. There could be one or more libraries that provide this functionality, such as \texttt{numpy}, \texttt{pandas}, RAPIDS CuDF \cite{cudf}, Acero (Apache Arrow Compute), etc, or locally developed as a part of \cylon{}. The choice of the library depends on the language runtime, the underlying memory format, and the hardware architecture. This is to prevent redundant development efforts for reinventing the existing functionality.

    \subsection{Auxiliary Sub-operators}
    
    \textit{Partition} operators are essential for distributed memory applications. Partitioning determines how a local data partition is split into subsets so they can be sent across the network. This operator is closely tied with \textit{Shuffle} communication routine. Hash partition, range partition, and rebalance are several key auxiliary operators.

    \subsection{Parallel Processing Patterns \& Operator Implementations}
    
\begin{table*}[hptb]
    \centering
\begin{tabular}{|l|c|c|c|} 
\hline
\multicolumn{1}{|c|}{\textbf{Pattern}}                                                                         & \textbf{Operators}                                                              & \begin{tabular}[c]{@{}c@{}}\textbf{Result }\\\textbf{Semantic}\end{tabular} & \textbf{Communication}                                                       \\ 
\hline
\textbf{Embarrassingly parallel}                                                                                        & \begin{tabular}[c]{@{}c@{}}Select, Project, Map, \\Row-Aggregation\end{tabular} & Partitioned                                                                 & -                                                                            \\ 
\hline
\textbf{Loosely Synchronous}                                                                                            & \multicolumn{1}{l|}{}                                                           & \multicolumn{1}{l|}{}                                                       & \multicolumn{1}{l|}{}                                                        \\
\begin{tabular}{@{\labelitemi\hspace{\dimexpr\labelsep+0.5\tabcolsep}}l@{}}Shuffle Compute\end{tabular}        & \begin{tabular}[c]{@{}c@{}}Union, Difference, \\Join, Transpose\end{tabular}    & Partitioned                                                                 & Shuffle                                                                      \\
\begin{tabular}{@{\labelitemi\hspace{\dimexpr\labelsep+0.5\tabcolsep}}l@{}}Combine Shuffle Reduce\end{tabular} & Unique, GroupBy                                                                 & Partitioned                                                                 & Shuffle                                                                      \\
\begin{tabular}{@{\labelitemi\hspace{\dimexpr\labelsep+0.5\tabcolsep}}l@{}}Broadcast Compute\end{tabular}      & Broadcast-Join$^{*}$                                                            & Partitioned                                                                 & Bcast                                                                        \\
\begin{tabular}{@{\labelitemi\hspace{\dimexpr\labelsep+0.5\tabcolsep}}l@{}}Globally Reduce\end{tabular}        & Column-Aggregation                                                              & Replicated                                                                  & AllReduce                                                                    \\
\begin{tabular}{@{\labelitemi\hspace{\dimexpr\labelsep+0.5\tabcolsep}}l@{}}Sample Shuffle Compute\end{tabular}       & Sort                                                                            & Partitioned                                                                 & \begin{tabular}[c]{@{}c@{}}Gather, Bcast, Shuffle, AllReduce\end{tabular}  \\
\begin{tabular}{@{\labelitemi\hspace{\dimexpr\labelsep+0.5\tabcolsep}}l@{}}Halo Exchange\end{tabular}          & Window                                                                          & Partitioned                                                                 & Send-recv                                                                    \\ 
\hline
\textbf{Partitioned I/O}                                                                                          & Read/Write                                                                      & Partitioned                                                                 & \begin{tabular}[c]{@{}c@{}}Send-recv, Scatter, Gather\end{tabular}         \\ 
\hline
\multicolumn{4}{c}{*Specialized join algorithm}          
\end{tabular}
    \caption{Generic Dataframe Operator Patterns}
    \label{tab:op-patterns}
\end{table*}

    According to our previous publication, dataframe operators can be broadly separated into three categories \cite{perera2022high}, as described in Table \ref{tab:op-patterns}.
    \begin{enumerate}
        \item Embarrassingly parallel: Operators that require no communication required 
        \item Loosely synchronous: Operators that require communication at some stage in its implementation. This is a broad category; therefore, it is separated into the following subcategories. 
                \begin{enumerate}
                    \item Shuffle-compute 
                    \item Sample-shuffle-compute
                    \item Combine-shuffle-reduce
                    \item Broadcast-compute 
                    \item Globally reduce
                    \item Halo exchange
                \end{enumerate}
        \item Partitioned I/O: I/O operators in distributed memory parallel environments require communication to load balance data amongst the workers. 
    \end{enumerate}

\section{Cost Model For Evaluation}\label{sec:cost-model}
    
A cost model can be applied to the \cylon{} distributed operator model to estimate the execution time/ cost of each operator pattern. As observed before, each pattern comprises three sub-operators. Hence, the total cost estimate ($T_{total}$) is the sum of the cost of each sub-operator. 

\vspace{1em}
\begin{minipage}{\linewidth}
    \begin{minipage}[c]{\textwidth}
    \centering
    \Large
    $T_{total} = T_{core} + T_{aux} + T_{comm}$
    \end{minipage}\vspace{1em}
    \begin{minipage}[c]{\textwidth}
    \centering
        \begin{enumerate}
            \itemsep0em 
            \item $T_{core} \rightarrow$ Core local operator cost
            \item $T_{aux} \rightarrow$ Auxiliary local operator cost
            \item $T_{comm} \rightarrow$ Communication operator cost
        \end{enumerate}
    \end{minipage}
\end{minipage}
\vspace{1em}

We analyze the communication and computation cost of distributed dataframe operators in the subsequent sections, and the following notation has been used. 
    \begin{itemize}
        \itemsep0em
        \item $P \rightarrow$ Parallelism
        \item $N \rightarrow$ Total number of rows 
        \item $n = N/P \rightarrow$ Number of rows per process
        \item $c \rightarrow$ Number of columns (constant for row-partitioned data)
        \item $\mathbf{N} = N\times c \rightarrow$ Total amount of distributed work/ total data
        \item $\mathbf{n} = \mathbf{N}/P \rightarrow$ Work per process/ rows per process
        \item $\mathbf{C} \rightarrow$ Cardinality of data
    \end{itemize}
    
    \subsection{Communication Cost ($T_{comm}$)}
    
    Based on the literature, Hockney \cite{hockney1994communication}, LogP \cite{culler1993logp}, and LogGP \cite{alexandrov1997loggp} are some of the most commonly used cost models to evaluate collective communication operations. \textit{Hockney model} provides a simple communication cost estimation, and therefore, it has been used in many recent publications \cite{thakur2005optimization,traff2014implementing,bruck1997efficient,pjevsivac2007performance}. The model fails to capture the network congestion. However, it provides an adequate cost estimation to evaluate \cylon{}. The model assumes that the taken to send a message between any two nodes can be modeled as,
    
    \vspace{1em}
    \begin{minipage}{0.99\linewidth}
        \begin{minipage}[c]{0.29\linewidth}
        \centering
        \Large
        $T = \alpha + n\beta$ 
        \end{minipage}
    \hfill
        \begin{minipage}[c]{0.69\linewidth}
        \centering
            \begin{enumerate}
                \itemsep0em 
                \item $n \rightarrow$ Message size/ number of bytes transferred
                \item $\alpha \rightarrow$ Latency/ startup time per message (independent of $n$)
                \item $\beta \rightarrow$ Transfer time per byte
            \end{enumerate}
        \end{minipage}
    \end{minipage}
    \vspace{1em}

    Let us take \textit{Shuffle (AllToAll)} for an example. \cylon{} uses non-blocking send-receive-based implementation. Each worker would shuffle $\mathbf{n}$ data with others in $P$ iterations. In each iteration, it would send and receive $\frac{\mathbf{n}}{P}$ amount of data (on average, for uniformly distributed data). Out of the $P$ iterations, one iteration is a local data transfer. Therefore, 
    \begin{center}
    $T_{shuffle} = (P-1)(\alpha + \frac{\mathbf{n}}{P}\beta) = (P-1)\alpha + \frac{(P-1)\mathbf{n}}{P}\beta$ \\
    \end{center}
    Therefore, for row-partitioned data, \\
    \begin{center}
    $T_{shuffle} = T_{startup} + T_{transfer} = O(P) + O(\frac{P-1}{P}\times n)$    
    \end{center}
    Table \ref{tab:comm-comp} describes the communication costs of communication routines used in distributed dataframe operator implementations for multiple algorithms based on the Hockney model. It uses the definitions described in the Section \ref{sec:cost-model}.

    \begin{table*}[hptb]
    \centering
    \begin{tabular}{ccccc}
    \hline
    \multicolumn{1}{|c|}{\textbf{Operation}}                  & \multicolumn{1}{c|}{\textbf{Algorithm}}                                                                  & \multicolumn{1}{c|}{\textbf{\begin{tabular}[c]{@{}c@{}}Startup\\time\\$(T_{startup})$\end{tabular}}} & \multicolumn{1}{c|}{\textbf{\begin{tabular}[c]{@{}c@{}}Transfer\\Time\\$(T_{transfer})$\end{tabular}}} & \multicolumn{1}{c|}{\textbf{\begin{tabular}[c]{@{}c@{}}Reduction\\Time\\$(T_{reduce})$\end{tabular}}} \\ \hline
    \multicolumn{1}{|c|}{\multirow{4}{*}{Shuffle/AllToAll}} & \multicolumn{1}{l|}{isend-irecieve\cite{thakur2005optimization}}                                                                      & \multicolumn{1}{c|}{$O(P)$}                  & \multicolumn{1}{c|}{$O(\frac{P-1}{P}*n)$}                  & \multicolumn{1}{c|}{-}                       \\ \cline{2-5} 
    \multicolumn{1}{|c|}{}                                    & \multicolumn{1}{l|}{Ring\cite{traff2014implementing}}                                                                                & \multicolumn{1}{c|}{$O(P)$}                & \multicolumn{1}{c|}{$O(P*n$)}         & \multicolumn{1}{c|}{-}                       \\ \cline{2-5} 
    \multicolumn{1}{|c|}{}                                    & \multicolumn{1}{l|}{Pairwise Exchange\cite{thakur2005optimization}}                                                                   & \multicolumn{1}{c|}{$O(P)$}                & \multicolumn{1}{c|}{$O(n)$}                  & \multicolumn{1}{c|}{-}                       \\ \cline{2-5} 
    \multicolumn{1}{|c|}{}                                    & \multicolumn{1}{l|}{Bruck\cite{bruck1997efficient}/ Modified Bruck\cite{traff2014implementing}}                                                               & \multicolumn{1}{c|}{$O(\log_{}P)$}               & \multicolumn{1}{c|}{$O(\log_{}P*\frac{n}{2})$}          & \multicolumn{1}{c|}{-}                       \\ \hline
    \multicolumn{1}{|c|}{\multirow{3}{*}{AllGather}}          & \multicolumn{1}{l|}{Ring\cite{thakur2005optimization}}                                                                                & \multicolumn{1}{c|}{$O(P)$}                      & \multicolumn{1}{c|}{$O(\frac{P-1}{P}*N)$}                       & \multicolumn{1}{c|}{-}                       \\ \cline{2-5} 
    \multicolumn{1}{|c|}{}                                    & \multicolumn{1}{l|}{Recursive Doubling\cite{thakur2005optimization}} & \multicolumn{1}{c|}{$O(\log_{}P)$}                      & \multicolumn{1}{c|}{$O(\frac{P-1}{P}*N)$}                       & \multicolumn{1}{c|}{-}                       \\ \cline{2-5} 
    \multicolumn{1}{|c|}{}                                    & \multicolumn{1}{l|}{Bruck\cite{thakur2005optimization}}                                                                               & \multicolumn{1}{c|}{$O(\log_{}P)$}                      & \multicolumn{1}{c|}{$O(\frac{P-1}{P}*N)$}                       & \multicolumn{1}{c|}{-}                       \\ \hline
    \multicolumn{1}{|c|}{\multirow{2}{*}{Broadcast}}          & \multicolumn{1}{l|}{Binomial Tree\cite{thakur2005optimization}}                                                                       & \multicolumn{1}{c|}{$O(\log_{}P)$}                      & \multicolumn{1}{c|}{$O(\log_{}P*n)$}                       & \multicolumn{1}{c|}{-}                       \\ \cline{2-5} 
    \multicolumn{1}{|c|}{}                                    & \multicolumn{1}{l|}{Scatter-AllGather \cite{shroff2000collmark}}         & \multicolumn{1}{c|}{$O(\log_{}P+P)$}                   & \multicolumn{1}{c|}{$O(\frac{P-1}{P}*n)$}                       & \multicolumn{1}{c|}{-}                       \\ \hline
    \multicolumn{1}{|c|}{\multirow{2}{*}{Reduce}}             & \multicolumn{1}{l|}{Binomial Tree\cite{thakur2005optimization}}                                                                       & \multicolumn{1}{c|}{$O(\log_{}P)$}                      & \multicolumn{1}{c|}{$O(\log_{}P*n)$}                       & \multicolumn{1}{c|}{$O(\log_{}P*n)$}                        \\ \cline{2-5} 
    \multicolumn{1}{|c|}{}                                    & \multicolumn{1}{l|}{Reduce-Scatter Gather \cite{rabenseifner2004optimization}}     & \multicolumn{1}{c|}{$O(\log_{}P)$}                      & \multicolumn{1}{c|}{$O(\frac{P-1}{P}*n)$}                       & \multicolumn{1}{c|}{$O(\frac{P-1}{P}*n)$}                        \\ \hline
    \multicolumn{1}{|c|}{\multirow{3}{*}{AllReduce}}          & \multicolumn{1}{l|}{Binomial Tree\cite{thakur2005optimization}}                                                                       & \multicolumn{1}{c|}{$O(\log_{}P)$}                      & \multicolumn{1}{c|}{$O(\log_{}P*n)$}                       & \multicolumn{1}{c|}{$O(\log_{}P*n)$}                        \\ \cline{2-5} 
    \multicolumn{1}{|c|}{}                                    & \multicolumn{1}{l|}{Recursive Doubling\cite{thakur2005optimization}}                                                                  & \multicolumn{1}{c|}{$O(\log_{}P)$}                      & \multicolumn{1}{c|}{$O(\log_{}P*n)$}                       & \multicolumn{1}{c|}{$O(\log_{}P*n)$}                        \\ \cline{2-5} 
    \multicolumn{1}{|c|}{}                                    & \multicolumn{1}{l|}{Reduce-Scatter AllGather \cite{rabenseifner2004optimization}}  & \multicolumn{1}{c|}{$O(\log_{}P)$}                      & \multicolumn{1}{c|}{$O(\frac{P-1}{P}*n)$}                       & \multicolumn{1}{c|}{$O(\frac{P-1}{P}*n)$}                        \\ \hline
    \end{tabular}
    \caption{Complexity of Communication Operations}
    \label{tab:comm-comp}
    \end{table*}

    \subsection{Computation Cost ($T_{core} + T_{aux}$)}
    
    Core local operator cost ($T_{core}$) \& auxiliary local operator cost ($T_{aux}$) constitutes the computation cost. Since these are local operations, the cost can be derived from \textit{time complexity of the algorithm}. 
    For example, a local \texttt{sort} operation would take (when using a quick-sort algorithm for uniformly distributed data), 
    \begin{center}
    $T_{sort} = O(n \log_{}{n})$ 
    \end{center}
    Table \ref{tab:com-complex} describes the time complexities of commonly used local dataframe operators (Core local operator cost, $T_{core}$) and their output size $(n_{new})$. 
        
\begin{table*}[hptb]
\centering
\begin{tabular}{|c|c|c|}
\hline
\multicolumn{1}{|c|}{\textbf{Local Operation}} & \textbf{Cost ($T_{core}$)}               & \textbf{Output Size $(n_{new})$} \\ \hline
Selection, Map                                & $O(n)$                                   & $O(n)$                           \\ \hline
Row-aggregation                               & $O(nc)=O(n)$                             & $O(n)$                           \\ \hline
Projection                                    & $O(c)$                                   & $O(nc)$                          \\ \hline
Union                                         & $O(nc) = O(n)$ (hash-based)              & $O(n\mathbf{C})$                 \\ \hline
Set-difference                                & $O(nc) = O(n)$ (hash-based)              & $O(n)$                           \\ \hline
Hash-Join                                     & $O(n)+ O(\frac{n}{\mathbf{C}})$          & $O(\frac{n}{\mathbf{C}})$        \\ \hline
Sort-Join                                     & $O(n\log_{}n) + O(\frac{n}{\mathbf{C}})$ & $O(\frac{n}{\mathbf{C}})$        \\ \hline
Transpose                                     & $O(nc)$                                  & $O(nc)$                          \\ \hline
Unique                                        & $O(nc) = O(n)$ (hash-based)              & $O(n\mathbf{C})$                 \\ \hline
GroupBy                                       & $O(n)$ (hash-based)              & $O(n\mathbf{C})$                 \\ \hline
Column Aggregation                            & $O(nc)=O(n)$                             & $O(c)$                           \\ \hline
Sort                                          & $O(n\log_{}n)$                           & $O(n)$                           \\ \hline
\end{tabular}
\caption{Core local operator cost ($T_{core}$)}
\label{tab:com-complex}
\end{table*}

    \subsection{Total Cost of Dataframe Operator Patterns}
    We will look at the total cost of each operator pattern in the following subsections. 
    
        \subsubsection{Embarrassingly Parallel}        
        This is the most trivial class of operators since they do not require any communication to parallelize the computation. \textit{Select}, \textit{Project}, \textit{Map}, and \textit{Row-Aggregation} fall under this pattern. 
        Arithmetic operations (ex: \texttt{add}, \texttt{mul}, etc.) are also good examples of this pattern. Embarrassingly parallel distributed operators can simply call the corresponding local operator, and therefore the cost estimation of this pattern is, 
        \begin{center}
        $T_{EP} = O(n)$ 
        \end{center}
        

        \subsubsection{Shuffle Compute}
        
        This common pattern can be used for operators that depend on \textit{Equality/Key Equality of rows}. Of the core dataframe operators, \texttt{join, union} and \texttt{difference} directly fall under this pattern. In contrast, \texttt{transpose} follows a more nuanced approach. 
        
        Partitioning and shuffling communication routines rearrange the data so that equal/key-equal rows are on the same partition at the end of the operation. This guarantees that the corresponding local operation can be called at the end of the shuffling stage. \textit{Join, Union} and \textit{Difference} operators follow this pattern: 
        \begin{center}
        \fcolorbox{black}{white}{Partition}$\rightarrow$\fcolorbox{black}{white}{Split}$\rightarrow$\fcolorbox{black}{lightgray}{Shuffle}$\rightarrow$\fcolorbox{black}{white}{LocalOp}
        \end{center}
        Therefore, the cost estimation of shuffle compute for each worker is, 
        \begin{center}
        $T_{shuffle\_compute(hash)} = O(n) + O(P) + O(\frac{P-1}{P}\times n) + T_{core}$ \\
        $T_{shuffle\_compute(range)} = O(\log_{}{P}) + O(n) + O(P) + O(\frac{P-1}{P}\times n) + T_{core}$
        \end{center}
    
        Typically partitioning schemes (hash, range, etc.) are map operators and, therefore, access memory locations contiguously. These can be efficiently executed on modern SIMD-enabled hardware. However, the local operator may need to access memory randomly (e.g., a join that uses a hash table). Therefore, allowing the local operator to work on in-cache data improves the efficiency of the computation. This can be achieved by simply attaching a \textit{local partition} block at the end of the \texttt{shuffle}. 
        \begin{center}
        \fcolorbox{black}{white}{Partition}$\rightarrow$\fcolorbox{black}{white}{Split}$\rightarrow$\fcolorbox{black}{lightgray}{Shuffle}$\rightarrow$\fcolorbox{black}{white}{Partition}$\rightarrow$\fcolorbox{black}{white}{Split}$\rightarrow$\fcolorbox{black}{white}{LocalOp}
        \end{center}
        A more complex scheme would be to partition data into much smaller sub-partitions from the beginning of the pipeline. Possible gains on each scheme depend heavily on runtime characteristics such as the data distribution.

        \subsubsection{Sample Shuffle Compute}
        
        This pattern is an extension of the shuffle-compute pattern. Sampling is commonly used for operators such as distributed \texttt{sort}. It gives an overview of the data distribution, which needs to be communicated among the other workers to determine an ordered (range) partition scheme. This can be achieved trivially by calling \texttt{all reduce} operation, or by a composite of communication \& computation steps (eg. sample sort). 
        \begin{center}
        \fcolorbox{black}{white}{Sample}$\rightarrow$
        \fcolorbox{black}{lightgray}{Communicate insights}$\rightarrow$
        \fcolorbox{black}{white}{Partition}$\rightarrow$
        \fcolorbox{black}{white}{Split}$\rightarrow$
        \fcolorbox{black}{lightgray}{Shuffle}$\rightarrow$
        \fcolorbox{black}{white}{LocalOp}
        \end{center}
        \cylon{} uses multiple algorithms for distributed \texttt{sort} implementation. The data can be range-partitioned for numerical key columns based on a key-data histogram,  and it would have the following total cost per worker. 
        \begin{center}
        \fcolorbox{black}{white}{Sample}$\rightarrow$
        \fcolorbox{black}{lightgray}{Allreduce range}$\rightarrow$
        \fcolorbox{black}{white}{Binning \&Range part.}$\rightarrow$
        \fcolorbox{black}{lightgray}{Shuffle}$\rightarrow$
        \fcolorbox{black}{white}{Local sort}
        \end{center}
        \begin{center}
        $T_{sort(range)} = O(\log_{}{P}) + O(n) + O(P) + O(\frac{P-1}{P}\times n) + O(n\log_{}{n})$ 
        \end{center}
            
        For the rest, \cylon{} uses \textit{sample sort} with regular sampling \cite{li1993versatility}. It sorts data locally and sends a sample to a central entity that determines pivot points for data. Based on these points, sorted data will be split and shuffled. Finally, all executors merge the received sub-partitions locally. 
        
        \begin{center}
        \fcolorbox{black}{white}{Local sort}$\rightarrow$
        \fcolorbox{black}{white}{Sample}$\rightarrow$
        \fcolorbox{black}{lightgray}{Gather @rank0}$\rightarrow$
        \fcolorbox{black}{white}{Calc. pivots @rank0}$\rightarrow$
        \fcolorbox{black}{lightgray}{Bcast pivots}$\rightarrow$
        \fcolorbox{black}{white}{Split}$\rightarrow$
        \fcolorbox{black}{lightgray}{Shuffle}$\rightarrow$
        \fcolorbox{black}{white}{Local merge}
        \end{center}

        \subsubsection{Combine Shuffle Reduce}
        
        Another extension of the \textit{Shuffle-Compute} pattern, Combine-Shuffle-Reduce, is semantically similar to the map-reduce \cite{dean2008mapreduce} paradigm. The operations that reduce the output length, such as \textit{Groupby} and \textit{Unique}, benefit from this pattern. The effectiveness of combine-shuffle-reduce over shuffle-compute depends on the \textit{Cardinality} $(\mathbf{C})$ (i.e., the ratio of unique rows to the total length). It follows, 
        \begin{center}
        \fcolorbox{black}{white}{LocalOp (interm. results)}$\rightarrow$
        \fcolorbox{black}{white}{Partition}$\rightarrow$
        \fcolorbox{black}{white}{Split}$\rightarrow$
        \fcolorbox{black}{lightgray}{Shuffle}$\rightarrow$
        \fcolorbox{black}{white}{LocalOp with final res.}
        \end{center}
        
        The initial local operation reduces data into a set of intermediate results (similar to the Combine step in \textit{MapReduce}), which would then be shuffled. Upon their receipt, a local operation is performed to finalize the results. The author also discusses this approach for dataframe reductions in a recent publication \cite{perera2020fast}.
        At the end of the initial local operation, the output dataframe size (in each worker) is $O(n\mathbf{C})$. Therefore, the total cost per worker would be, 
        \begin{center}
        $T_{comb\_shuf\_red} = T_{core}(n) + O(n\mathbf{C}) + O(P) + O(\frac{P-1}{P}\times n\mathbf{C}) + T_{core}(n\mathbf{C})$
        \end{center}

            
            
        \subsubsection{Globally Reduce}
        
        This pattern is most commonly seen in dataframe \textit{Column-Aggregation} operators. It is similar to the embarrassingly parallel pattern but requires an extra communication step to arrive at the final result. For example, calculating the column-wise \texttt{mean} requires a local summation, a global reduction, and a final value calculation. 
        \begin{center}
        \fcolorbox{black}{white}{LocalOp}$\rightarrow$
        \fcolorbox{black}{lightgray}{Allreduce}$\rightarrow$
        \fcolorbox{black}{white}{Finalize}
        \end{center}
        Some utility methods such as \textit{distributed length} and \textit{equality} also follow this pattern. For large data sets, the complexity of this operator is usually governed by the computation rather than the communication.
            
        \subsubsection{Halo Exchange}
        
        This pattern is observed in window operations. A window operation performs an aggregation over a sliding partition of values. Pandas API supports rolling and expanding windows. For row partitions, the windows at the boundaries would have to communicate with their neighboring partitions and exchange partially computed results. The amount of data sent/received is based on the window type and individual length of partitions.
        
        \subsubsection{Broadcast Compute}
        
        Broadcast compute is a scaled-down pattern from shuffle-compute. Rather than shuffling, certain operators like broadcast-join can use broadcasting. This strategy only becomes useful when there is a smaller relation so that it can be broadcasted without shuffling the large relation. It reduces communication overhead significantly. However, broadcast-joins would perform poorly if the relations were of the same order. This effect was observed in Modin \cite{petersohn2020towards}, where out-of-memory errors are reported even for moderately large datasets because it only employs broadcast joins.

        \begin{center}
        \fcolorbox{black}{lightgray}{Broadcast}$\rightarrow$\fcolorbox{black}{white}{LocalOp}
        \end{center}
        

        \subsubsection{Partitioned I/O}
        
        \textit{Partitioned Input} parallelizes the input data (CSV, JSON, Parquet) by distributing the files to each executor. It may distribute a list of input files to each worker evenly. Alternatively, it receives a custom one-to-many mapping from the worker to input file(s). It reads the input files according to the custom assignment. For Parquet files, \textit{Partitioned Input} tries to distribute the number of rows to each partition as evenly as possible when metadata is present. Suppose an executor does not receive data from reading. In that case, it constructs an empty dataframe with the same schema as the other partitions. In \textit{Partitioned Output}, each executor writes its partition dataframe to one file.

    \subsection{Runtime Aspects}
    \label{sec:run_asp}
    
        \subsubsection{Cardinality} \label{sec:card}
        Equality of rows governs the \textit{Cardinality} of a Dataframe $\mathbf{C}$, which is the number of unique rows relative to the length. Therefore, $\mathbf{C} \in [\frac{1}{N}, 1]$, where $\mathbf{C}=\frac{1}{N} \implies$ rows are identical and $\mathbf{C}=1 \implies$ all rows are unique. 
        In the \textit{Combine-Shuffle-Reduce} pattern, the initial local operation has the potential to reduce communication order to $n^{\prime} < n$. This gain depends on the \textit{Cardinality} ($\mathbf{C}$) of the dataframe $\mathbf{C} \in [\frac{1}{N}, 1]$, which is the number of unique rows relative to the length. $\mathbf{C} \sim \frac{1}{N} \implies n^{\prime} \lll n$, making the combine-shuffle-reduce much more efficient than a shuffle-compute. Consequently, when $\mathbf{C} \sim 1 \implies n^{\prime} \sim n$ may in fact worsen the combine-shuffle-reduce complexity. In such cases, the shuffle-compute pattern is more efficient. This incident is very evident from the cost model. 
        \begin{center}
        $T_{comb\_shuf\_red} = T_{core}(n) + O(n\mathbf{C}) + O(P) + O(\frac{P-1}{P}\times n\mathbf{C}) + T_{core}(n\mathbf{C})$ \\
        vs \\
        $T_{shuf\_comp} = O(n) + O(P) + O(\frac{P-1}{P}\times n) + T_{core}(n)$
        \end{center}
        When, $\mathbf{C} \to 1 \implies T_{comb\_shuf\_red} \to T_{shuf\_comp}$, and in fact, it is worse because the core local operation would have to be carried out twice.       
        
        \subsubsection{Data Distribution}
        Data distribution heavily impacts the partitioning operators. Some executors may be underutilized when unbalanced partitions exist, affecting the overall distributed performance. \textit{Work-stealing} scheduling is a possible solution to this problem. In a BSP environment, pseudo-work-stealing execution can be achieved by storing partition data in a shared object-store. Furthermore, some operations could employ different operator patterns based on the data distribution. For instance, when one relation is very small by comparison, \textit{Join} could use a \texttt{broadcast\_join} (broadcast-compute) rather than a hash-shuffle join (shuffle-compute) to achieve better performance.


        \subsubsection{Out-of-Core Execution}
        Currently, \cylon{} is limited by the memory available to the workers. With the data immutability guarantees, it always allocates new memory for the columns that get modified. Therefore, loosely synchronous patterns may require a workspace of $3-4\times$ the size of the table. This could be a challenging requirement for memory-constrained environments and limits the dataset size we could process. Therefore, the system needs to be able to execute operators out-of-core. 

        \subsubsection{Logical Plan Optimizations}
        
        A typical SQL query may translate to multiple Dataframe operators, and the application script can include several such queries. Semantically, these operators construct a DAG (directed acyclic graph) or a \textit{logical plan}. SQL and data engineering engines generate an \textit{optimized logical plan} based on rules (ex: predicate push-down) or cost metrics. While these optimizations produce significant gains in real-life applications, this is an orthogonal detail to the individual operator patterns we focus on in this paper.

\section{Experiments} \label{sec:experiments}

To evaluate the performance of \cylon{} distributed-memory execution model, we have conducted the following experiments. 
\begin{itemize}
    \item Communication and computation breakdown of \cylon{} operators for strong and weak scaling
    \item Running \cylon{} in Oak Ridge National Laboratory Summit supercomputer 
    \item Comparing \cylon{} performance against the state-of-the-art data processing systems
\end{itemize}

For the following experiments, uniformly random distributed data was used with two \texttt{int64} columns in column-major format (Fortran order). Data uses a cardinality of 90\% (i.e. 90\% of rows are unique), which constitutes a worst-case scenario for key-based operators (eg. join, sort, groupby, etc). The main focus of these experiments is to micro-benchmark the distributed operator implementation. Using a generated dataset allows the input dataset to be uniformly distributed and thereby evaluate the true performance of the kernels. Barthels et al. followed a similar approach to evaluate distributed join kernels \cite{barthels2017distributed}.

    \subsection{Communication \& Computation} \label{sec:comm-comp}

    These experiments were carried out on a 15-node Intel\textsuperscript{\textregistered} Xeon\textsuperscript{\textregistered} Platinum 8160 cluster. Each node comprises 48 hardware cores on two sockets, 255GB RAM, and SSD storage, and is connected via Infiniband with 40Gbps bandwidth. 

    \begin{figure}[htpb]
    \begin{center}
    \includegraphics[width=0.5\textwidth]{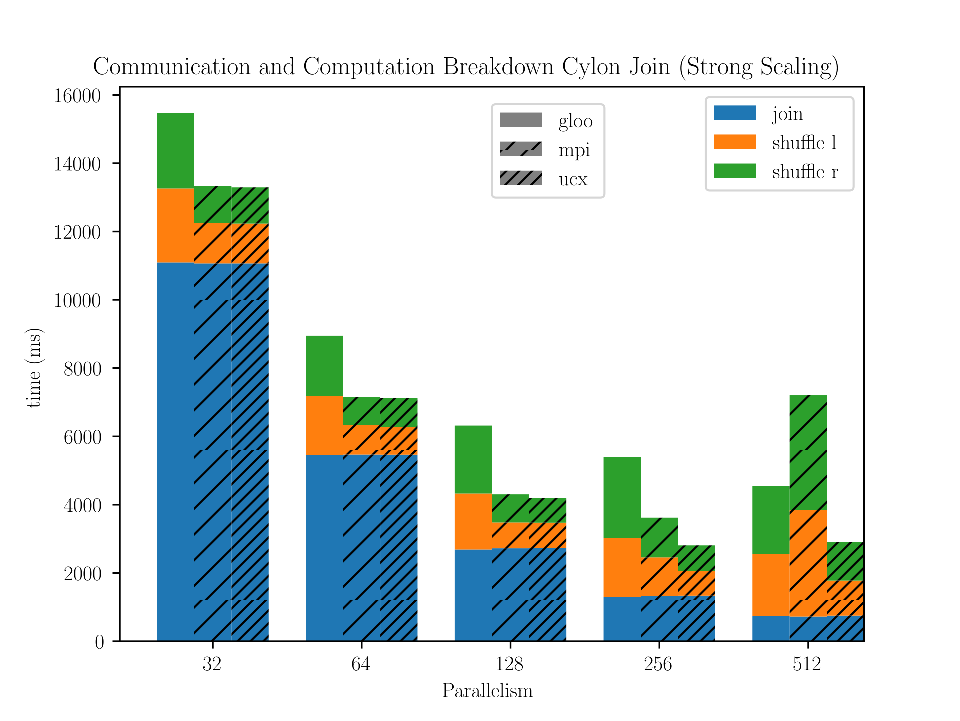}
    \end{center}
    \caption{Computation and Communication Breakdown - \texttt{join} (Strong Scaling)}
    \label{fig:comm-comp}
    \end{figure}

\begin{figure*}[htpb]
    \centering
    \begin{subfigure}{.5\linewidth}
        \centering
        \includegraphics[width=\linewidth]{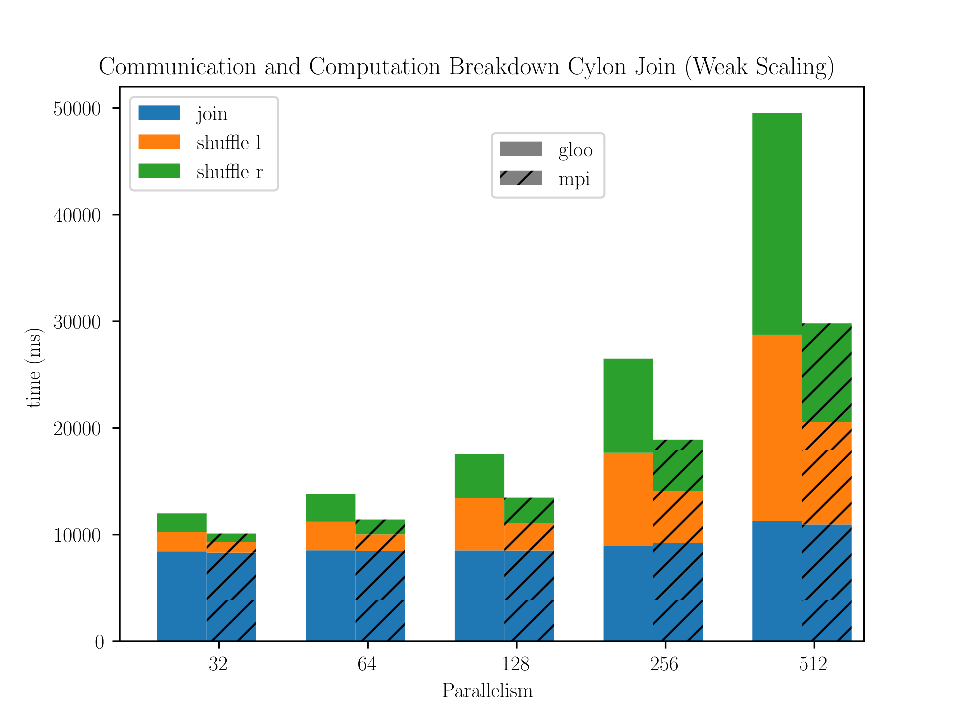}
    \end{subfigure}%
    \begin{subfigure}{.5\linewidth}
        \centering
        \includegraphics[width=\linewidth]{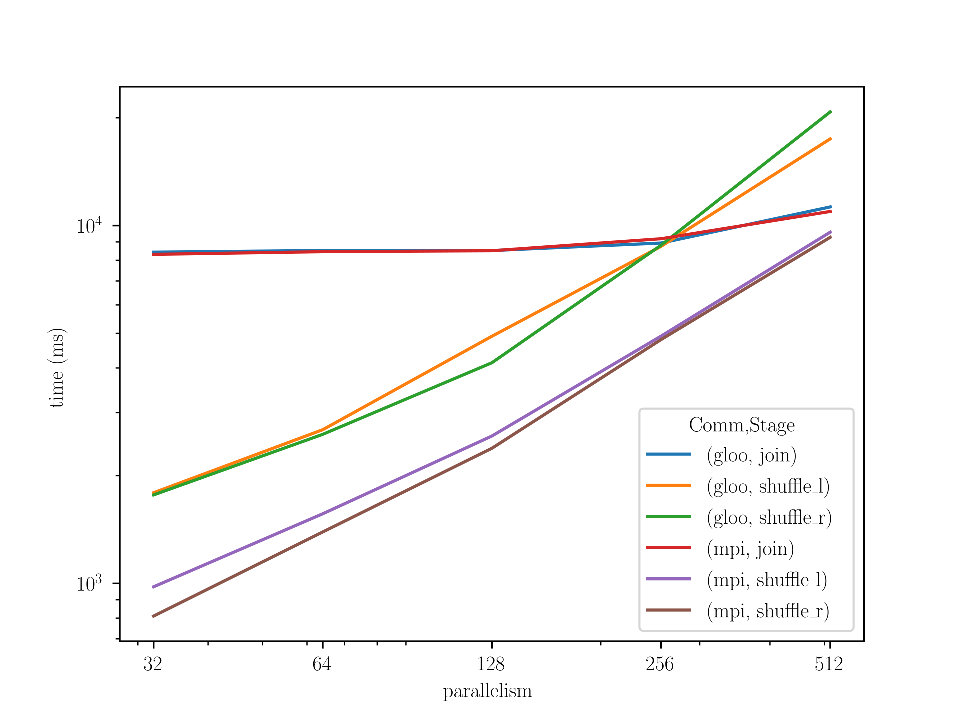}
    \end{subfigure}
    \caption{Computation and Communication Breakdown - \texttt{join} (Weak Scaling)}
    \label{fig:comm-comp-weak}
    \end{figure*}

    Figure \ref{fig:comm-comp} shows communication and computation time breakdown for \texttt{join} operation for a strong scaling test (1B rows per table). Moreover, Figure \ref{fig:comm-comp-weak} shows the same for a weak scaling test (25M per worker per table). Out of many operators, \texttt{join}s have the most communication overhead, as it is a binary operator (2 input DFs). 
    
    In the strong scaling plot, even at the smallest parallelism (32), there is a significant communication overhead (Gloo 27\%, MPI 17\%, UCX 17\%), and as the parallelism increases, it dominates the wall time (Gloo 76\%, MPI 86\%, UCX 69\%). Unfortunately, the author needed more expertise in the Spark, Dask, or Ray DDF code base to run a similar micro-benchmark. This experiment shows that communication plays a significant role in dataframe operator implementation. Despite using libraries specialized for message passing, \cylon{} still encounters significant communication overhead. Therefore, careful consideration must be given to communication while developing distributed dataframe runtimes.  
    
    The weak scaling plot can further analyze the impact of communication performance. The work per process is fixed; therefore, we should see a flat graph. However, as we see in Figure \ref{fig:comm-comp-weak}, the time increases along the parallelism axis, indicating that the communication overhead increases. The graph on the right plots each stage (log-log). The local join computation is relatively flat, while both shuffle stages (left \& right) show a linear increase. 
    
    \subsubsection{Examining the results using the cost model}
    
    By looking at the cost model in Section \ref{sec:cost-model}, the cost of \texttt{join} would be,
    \begin{center}
        $T_{shuffle} = O(P-1) + O(\frac{P-1}{P}\times n)$\\
        $T_{join(sort)} = O(P-1) + O(\frac{P-1}{P}\times n) + O(n) + O(n\log_{}n) + O(\frac{n}{\mathbf{C}})$ \\ 
        Substituting $n=N/P$, \\
        $T_{join(sort)} = O(P-1) + O(\frac{P-1}{P}\times \frac{N}{P}) + O(\frac{N}{P}) + O(\frac{N}{P}\log_{}{\frac{N}{P}}) + O(\frac{N}{P\mathbf{C}})$
    \end{center}
    
    For strong scaling, $N$ is constant. Therefore, as $P$ increases, the components that depend on $n$ (in computation and communication) reduce. This results in a downward trend in wall time. However, the $O(P-1)$ component (coming from the communication cost) overtakes the gains of reducing $n$. This explains the increase in wall time in higher parallelisms. 
    
    Similarly, for weak scaling, $n$ is kept constant, which reduces the cost to $O(P-1) + O(\frac{P-1}{P})$. For the parallelism values tested in the experiments (Figure \ref{fig:comm-comp-weak}), this explains the increasing wall-time values and linear upward trends in shuffle timings. Even though the amount of data transferred per worker remains constant ($n$), the cost model does not account for network congestion. This could explain the increasing gradient at higher parallelisms. 
    
    In the following sections, we will see that \cylon{} outperforms the state-of-the-art data engineering systems available today. However, the weak scaling indicates that \cylon{} still needs to improve on the communication operator performance (such as \textit{shuffle}). It would be worthwhile evaluating other algorithms such as Pairwise Exchange\cite{thakur2005optimization}, Bruck\cite{bruck1997efficient}/ Modified Bruck\cite{traff2014implementing}, etc., that have better time complexity as the parallelism increases. Another option would be to completely offload the shuffle implementation to the communication library (MPI, Gloo, UCX) and let the library decide which algorithm to choose based on runtime characteristics.

    \subsection{\cylon{} on ORNL Summit Supercomputer}

    \cylon{} was run on the Summit supercomputer at Oak Ridge National Laboratory (ORNL) as a part of large-scale testing. 
    \begin{figure}[htpb]
    \centering
    \includegraphics[width=\linewidth]{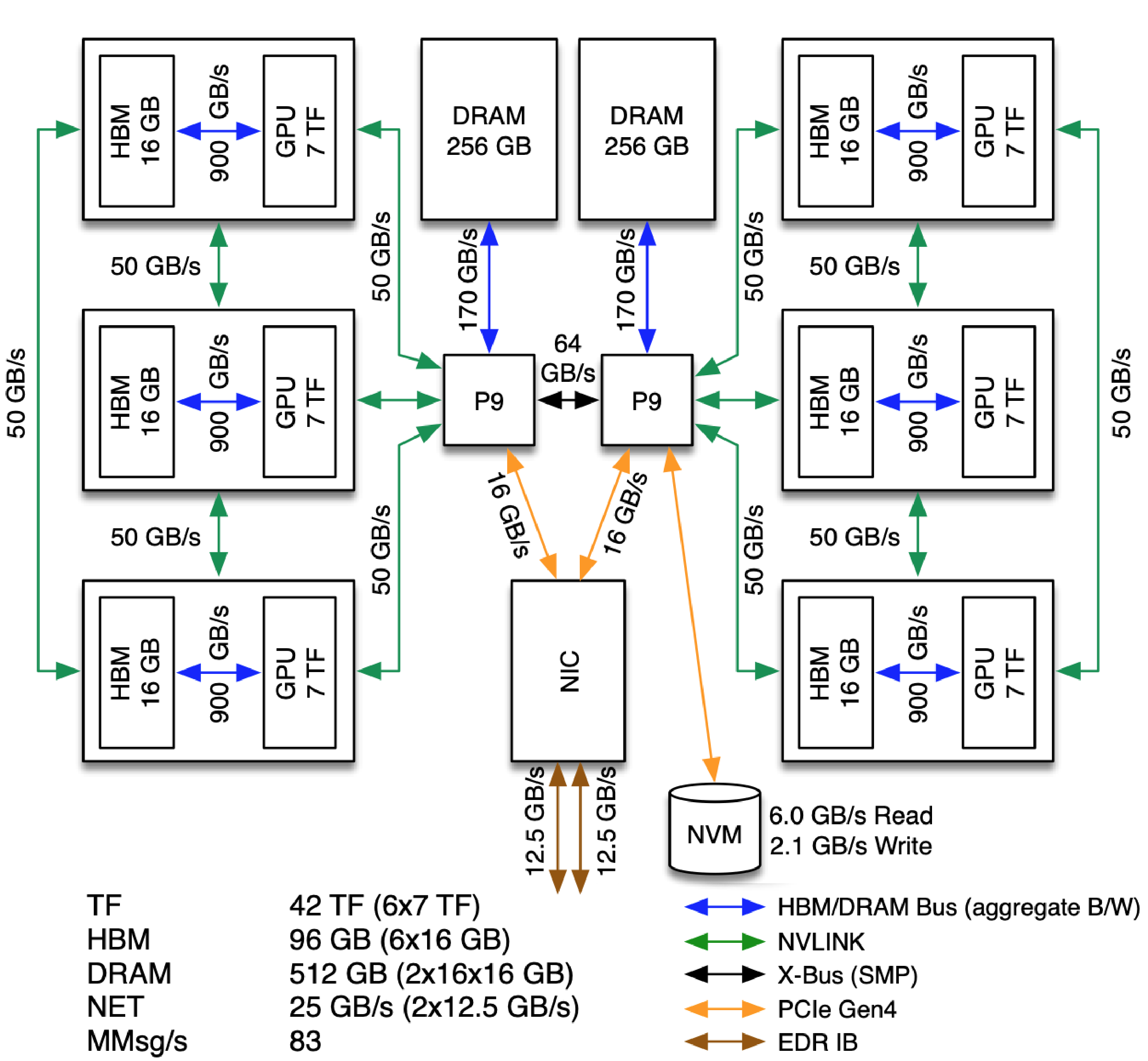} 
    \caption{ORNL Summit Node Architecture \cite{Summit89:online}}
    \label{fig:summit-arch}
    \end{figure}
    Each node in Summit consists of two IBM POWER9 processors and six Nvidia Tesla V100 accelerators, and there are 4600 of these nodes available for computation, reaching a theoretical peak double-precision performance of approximately 200 PF. Each node consists of 512 GB of RAM and 42 hardware cores. Figure \ref{fig:summit-arch} shows the architecture of a single node in Summit. For \cylon{} workloads, only the CPU nodes were used. 

    \subsubsection{Setting up \cylon{} in Summit}
    
    Setting up \cylon{} environment in Summit proved to be a tedious undertaking. Generally, \cylon{} is installed via a Conda Python environment \cite{Conda:online}, which conveniently installs dependencies using the official Anaconda packages. However, due to the Summit node hardware architecture, some of these default packages were failing unexpectedly. Most notably, we encountered memory allocation errors from the Apache Arrow library. Since this is an essential requirement for \cylon{}, we had to rebuild Apache Arrow natively on Summit hardware architecture. This was done by the native \cylon{} installation script which uses PyPI (\texttt{pip}) environment \cite{PyPI:online}.

    Additionally, Summit supercomputer uses its own MPI implementation based on IBM Spectrum MPI \cite{IBMSpect59:online}. At the time, \cylon{} was tested on OpenMPI and Microsoft MPI only, and therefore, several minor changes were required to properly link with Summit MPI modules. 

    The recommended way of using custom software in Summit is to create a module and load it (with dependencies) in batch scripts. However, this requires advanced expertise in Summit package management. We bypassed this requirement by installing \cylon{} and its dependencies into a PyPI environment using a login node. This PyPI environment resides in the user space in the file system. When submitting a batch job, we would activate this environment and run our \cylon{} script. 

    Following is an example batch script for a \cylon{} workload. 
    \begin{lstlisting}
    #!/bin/bash
    #BSUB -P <project name>
    #BSUB -W 1:30
    #BSUB -nnodes 8
    #BSUB -alloc_flags smt1
    #BSUB -J cylonrun-s-8
    #BSUB -o cylonrun-s-8.%J
    #BSUB -e cylonrun-s-8.%J
    
    module load python/3.7.7 gcc/9.3.0
    source $HOME/CYLON/bin/activate
    BUILD_PATH=$HOME/cylon/build
    export LD_LIBRARY_PATH=$BUILD_PATH/arrow/install/lib64:$BUILD_PATH/glog/install/lib64:$BUILD_PATH/lib64:$BUILD_PATH/lib:$LD_LIBRARY_PATH
    
    time jsrun -n $((8*42)) -c 1 python $HOME/cylon/summit/scripts/cylon_scaling.py -n 9999994368 -s s
    \end{lstlisting}  
    Both installation and batch scripts are available in the \cylon{} GitHub repository \cite{cylon}.

    \subsubsection{Strong Scaling}

\begin{figure*}[htpb]
    \centering
        \begin{subfigure}{.5\linewidth}
            \centering
            \includegraphics[width=\textwidth]{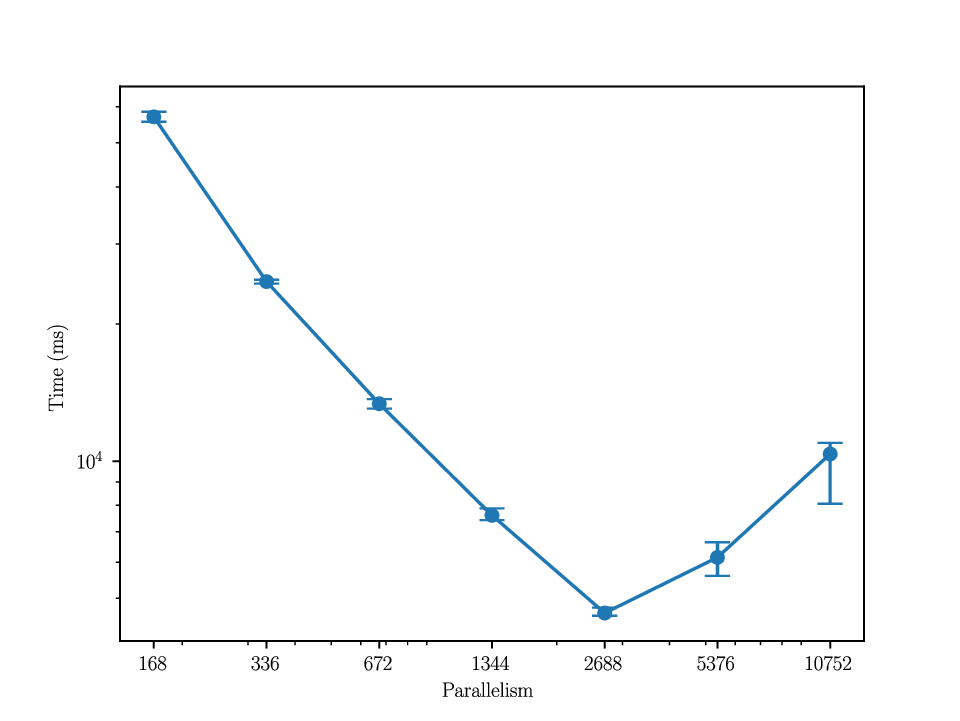}
            \caption{10B Rows}
            \label{fig:summit-strong10}
        \end{subfigure}%
        \begin{subfigure}{.5\linewidth}
            \centering
            \includegraphics[width=\textwidth]{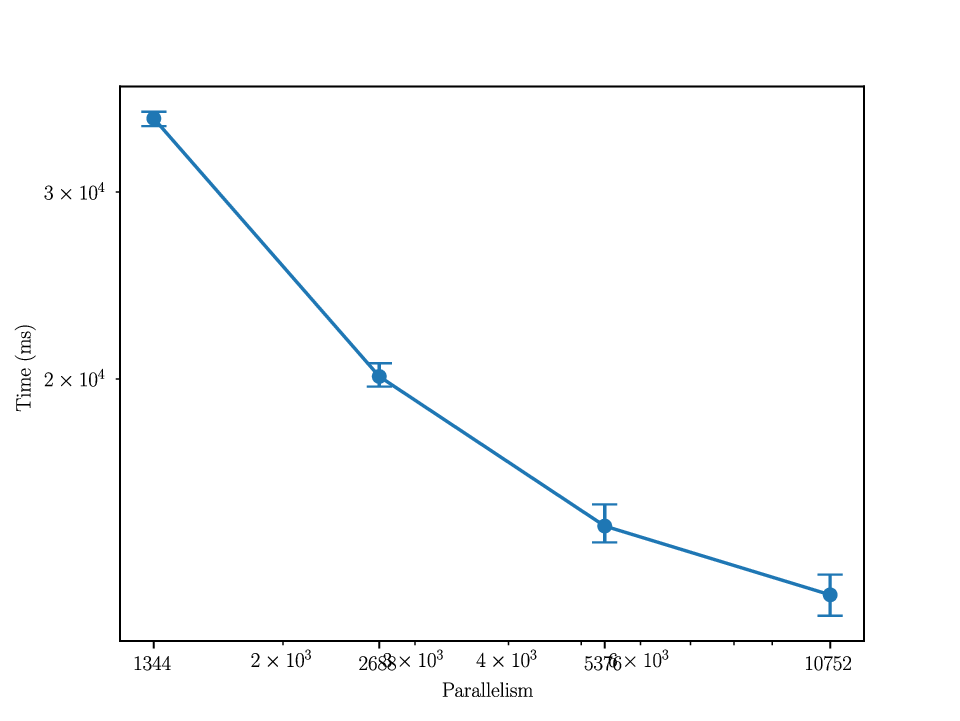}
            \caption{50B Rows}
            \label{fig:summit-strong50}
        \end{subfigure}
    
    \caption{\cylon{} Strong Scaling on Summit}
    \label{fig:summit-strong}
    \end{figure*}

    A strong scaling experiment was carried out on \cylon{} \texttt{join} operation of two 10 billion row tables. The size of each table is around 160GB. The parallelism was increased from 4 nodes ($4\times42=168$ cores) to 25 nodes ($256\times42=10,752$ cores). Figure \ref{fig:summit-strong} plots the results on a log-log scale. 
    
    Figure \ref{fig:summit-strong10} shows 10 billion rows per table experiment. As the parallelism increases from 168 to 2688, the wall time reduces almost linearly with fairly consistent timings. However, from thereon, the timings take a drastic turn and show a higher variance. From 5,376 onward, the computation component is less than 2 million rows per table per core. Therefore, communication would dominate the final wall time. 

    To further analyze this scenario, another 50 billion rows per table experiment was carried out (Figure \ref{fig:summit-strong50}). There, smaller parallelism experiments were unsuccessful due to memory limitations. However, for higher parallelisms, the wall time reduces fairly linearly, as expected. This indicates that, as long as the computation dominates the communication, performance gains can be achieved by adding more resources. For 50 billion cases, the inflection point would occur at higher parallelism than 10752.  
    
    \subsubsection{Weak Scaling}
    
\begin{figure}[htpb]
\centering
\includegraphics[width=\linewidth]{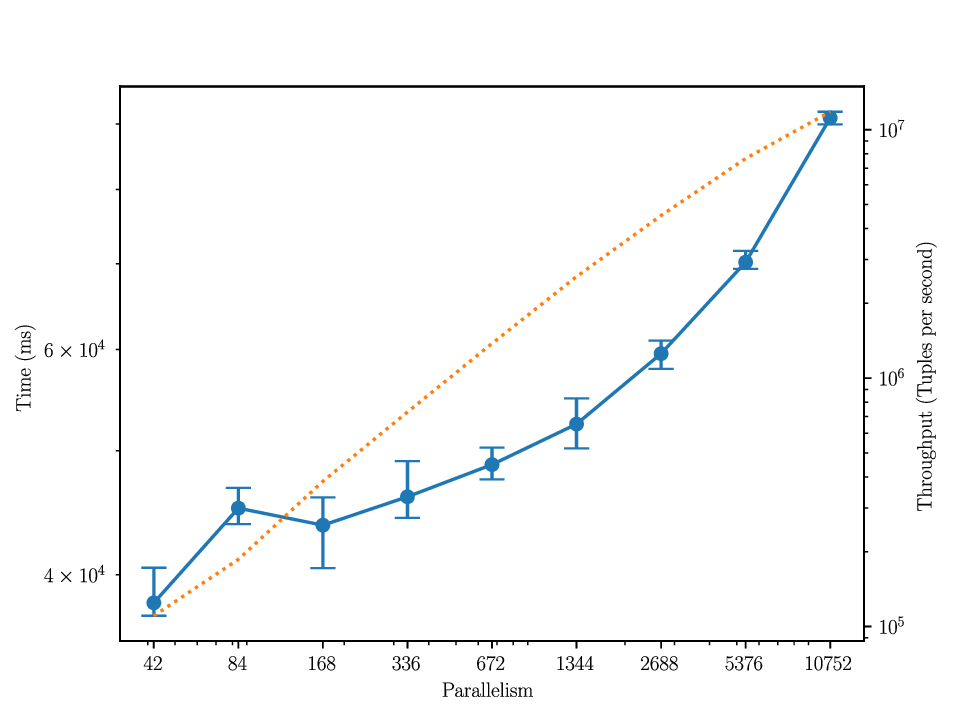}
\captionof{figure}{\cylon{} Weak Scaling on Summit}
\label{fig:summit-weak}
\end{figure}

    A weak scaling experiment was carried out again on \cylon{} \texttt{join} operation. The intention was to utilize the memory available in the node allocation fully. Considering the 512GB RAM and 42 cores per node, it was decided to use 50 million row tables per core. The number of cores has been increased from 1 to 10752, where the last experiment joins more than 1 trillion rows from the two tables. The results are depicted in Figure \ref{fig:summit-weak}.
    
    As we saw in the previous weak scaling experiments, the wall time increases with parallelism. This is not ideal for a weak scaling plot. However, the main culprit for this increase is the \textit{shuffle} communication overhead. However, \cylon{} was able to successfully process more than 17 terabytes (TB) of data across 10,752 cores which is a commendable achievement. When looking at the throughput of the operation, it steadily increases to close to 12 million tuples/second.

\begin{table}[htpb]
\centering
\scriptsize
    \begin{tabular}{|c|c|c|c|}
    \hline
    \textbf{Cores} & \textbf{\begin{tabular}[c]{@{}c@{}}Rows\\ (Mn)\end{tabular}} & \textbf{\begin{tabular}[c]{@{}c@{}}Size\\ (GB)\end{tabular}} & \textbf{\begin{tabular}[c]{@{}c@{}}Throughput\\ (Tuples/s)\end{tabular}} \\ \hline
    1              & 50                                                           & 1                                                            & 3,261                                                                    \\ \hline
    42             & 2,100                                                        & 34                                                           & 110,437                                                                  \\ \hline
    84             & 4,200                                                        & 67                                                           & 186,267                                                                  \\ \hline
    168            & 8,400                                                        & 134                                                          & 384,137                                                                  \\ \hline
    336            & 16,800                                                       & 269                                                          & 729,943                                                                  \\ \hline
    672            & 33,600                                                       & 538                                                          & 1,377,837                                                                \\ \hline
    1,344          & 67,200                                                       & 1,075                                                        & 2,561,797                                                                \\ \hline
    2,688          & 134,400                                                      & 2,150                                                        & 4,513,890                                                                \\ \hline
    5,376          & 268,800                                                      & 4,301                                                        & 7,657,451                                                                \\ \hline
    10,752         & 537,600                                                      & 8,602                                                        & 11,814,754                                                               \\ \hline
    \end{tabular}
\captionof{table}{Summit Weak Scaling Results}
\label{tab:my-table}        
\end{table}

    \subsection{\cylon{} vs. the State-of-the-art}

    In order to evaluate the performance of the distributed-memory execution model discussed in this paper, we performed a strong scaling analysis on several state-of-the-art distributed dataframe systems that are described in the related work section (Section \ref{sec:rel_work}). Experiments were also carried out on Pandas \cite{pandas:online} to get a serial performance baseline. The following frameworks were considered. We tried our best to refer to publicly available documentation, user guides, and forums while carrying out these tests to get the optimal configurations. 
    \begin{itemize}
        \item Dask Distributed Dataframes v2022.8
        \item Ray Datasets v1.12
        \item Modin Distributed Dataframe v0.13
        \item Apache Spark (Pandas-on-Spark) v3.3
    \end{itemize}
    
    We have carried out similar strong scaling analyses in the precursor publication \cite{perera2022high,perera2023high}, and several others \cite{perera2023towards, perera2023supercharging, widanage2020high}. In this publication, the results have been updated to the latest versions of software and their dependencies. The same 15-node Intel\textsuperscript{\textregistered} Xeon\textsuperscript{\textregistered} Platinum 8160 cluster described in Section \ref{sec:comm-comp} was also used for these experiments. 
    
    The following dataframe operator patterns were used for the experiments. When evaluating large-scale data engineering use cases (eg. TPC benchmarks \cite{TPCHomep1:online}, Deep Learning Recommendation Model 
    (DLRM) preprocessing \cite{Optimizi45:online}, etc) and based on our prior experience, these operator patterns \cite{widanage2020high,perera2023high} consume the majority of the computation time. 
    \begin{itemize}
        \item Shuffle Compute - \textit{Join} operator
        \item Combine Shuffle Reduce - \textit{GroupBy} operator
        \item Sample Shuffle Compute - \textit{Sort} operator
    \end{itemize}

\begin{figure*}[htbp]
    \centering
    \begin{tabular}{ccc}
        \includegraphics[width=0.49\textwidth]{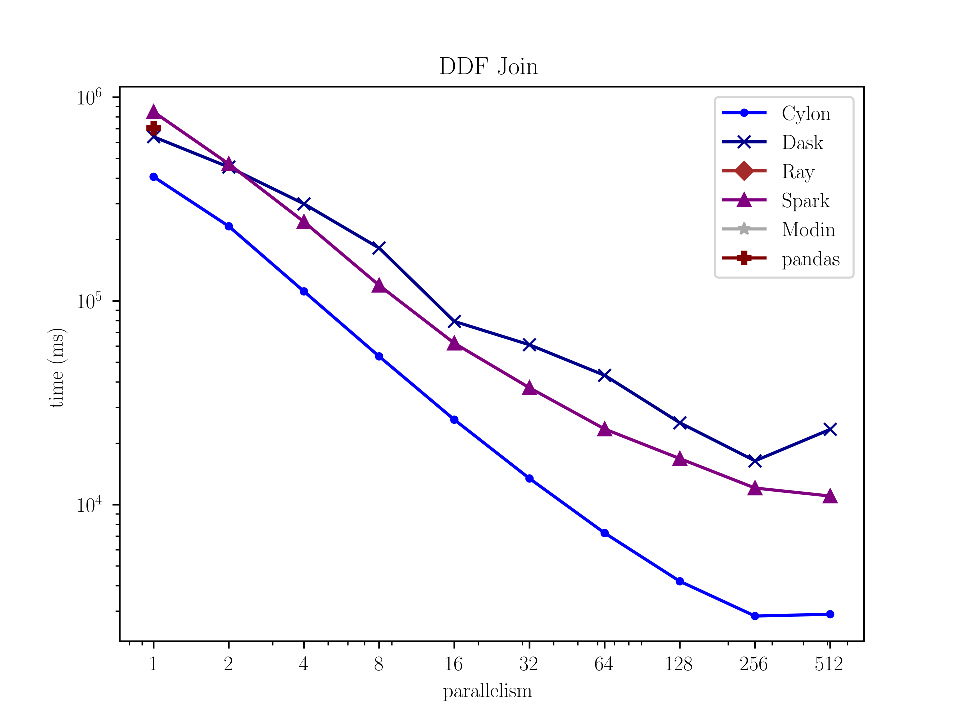} & 
        \includegraphics[width=0.49\textwidth]{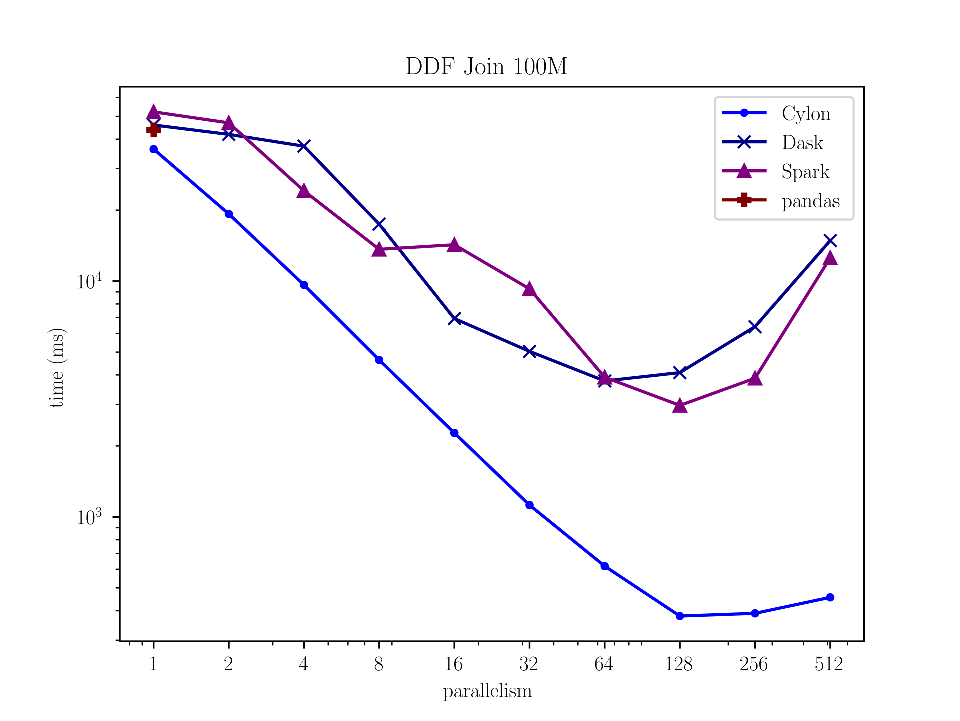} \\
        \includegraphics[width=0.49\textwidth]{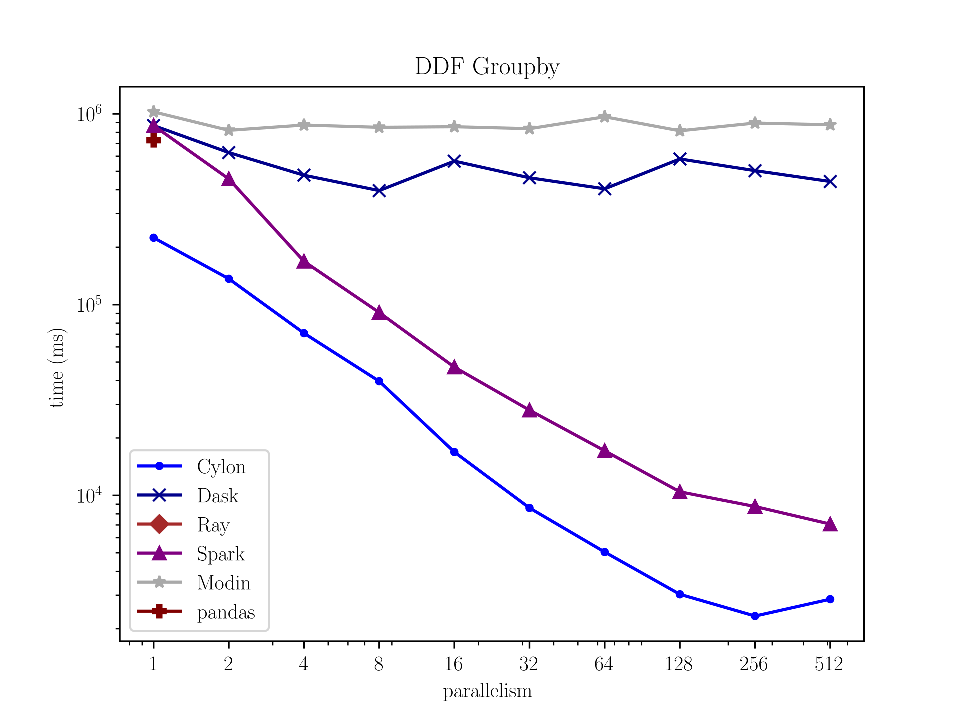} &
        \includegraphics[width=0.49\textwidth]{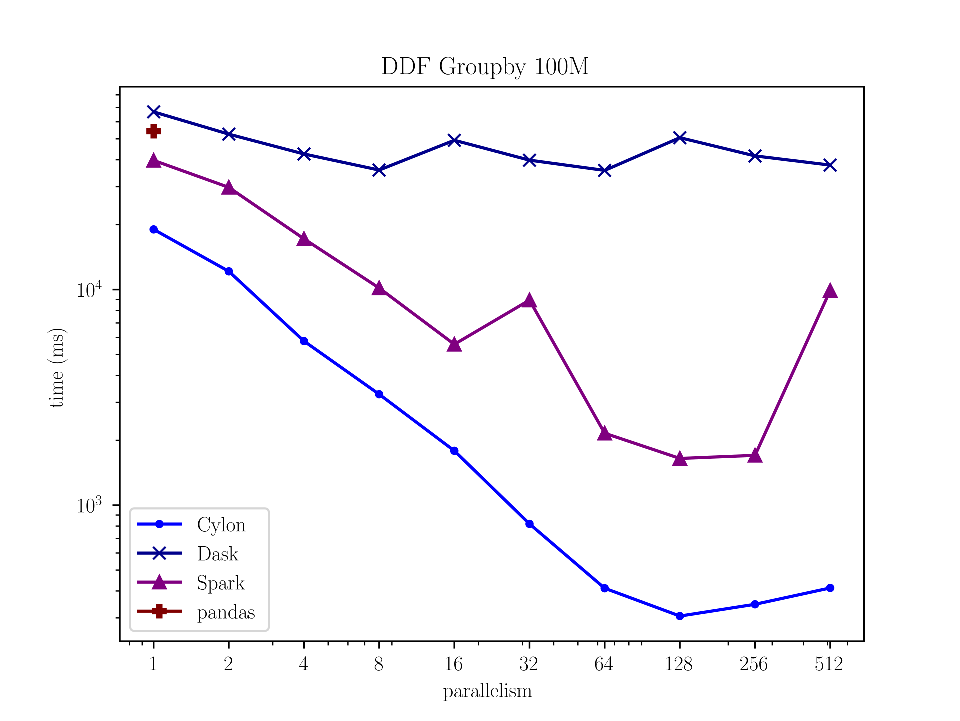} \\
        \includegraphics[width=0.49\textwidth]{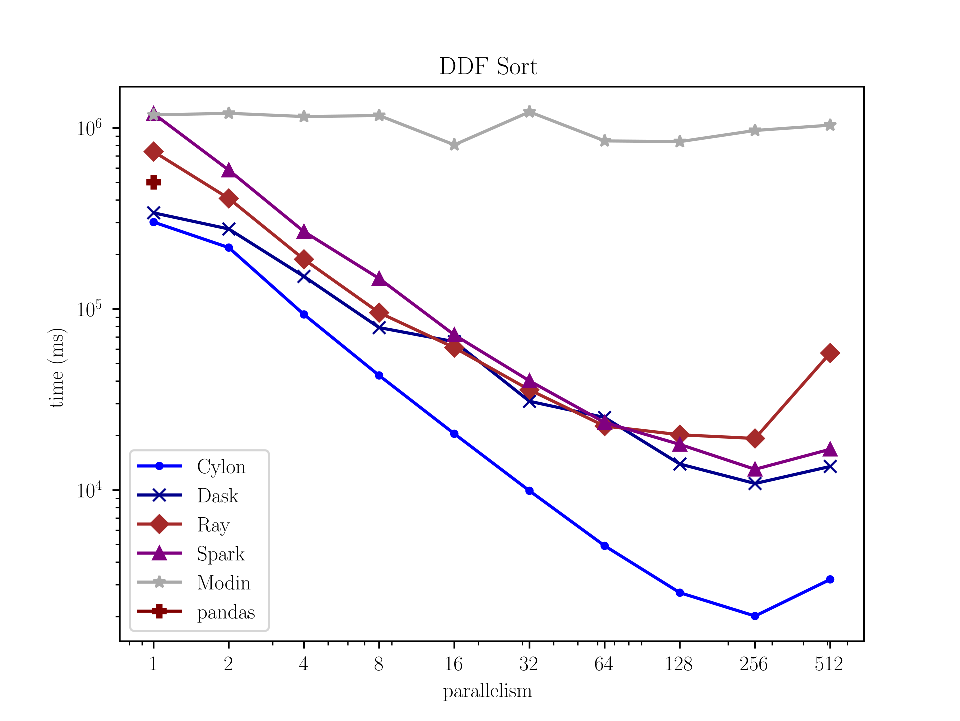} &
        \includegraphics[width=0.49\textwidth]{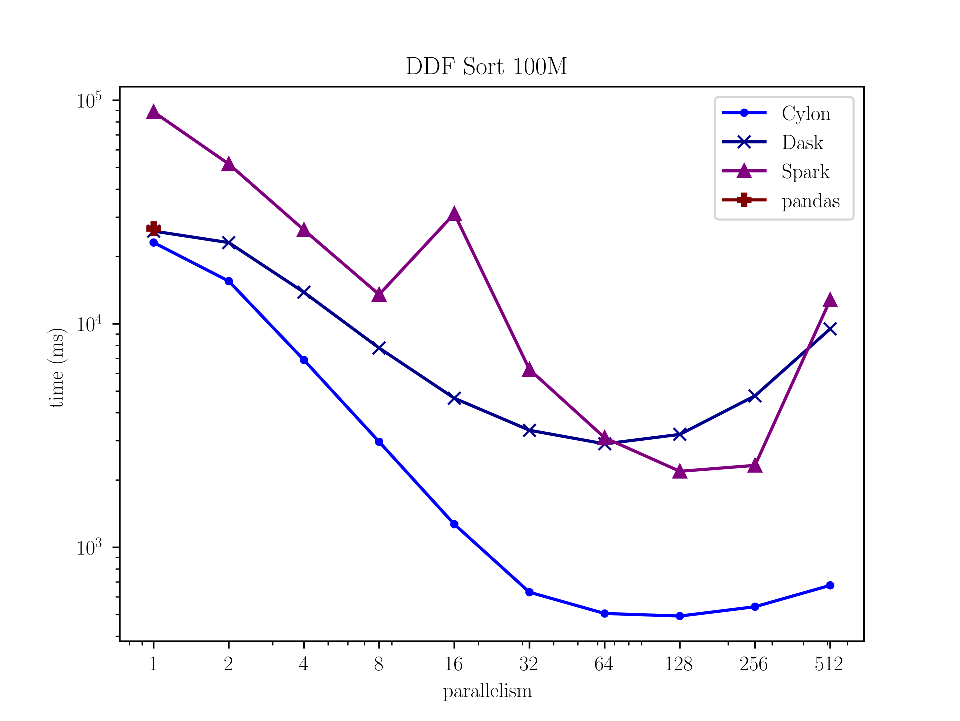} 
    \end{tabular} 
    \caption{Strong Scaling of Distributed Dataframe Operators (Log-Log), Left: 1B rows, Right: 100M rows (Only \cylon{}, Dask, \& Spark)}
    \label{fig:sperf}
    \end{figure*}

    Figure \ref{fig:sperf} depicts two sets of strong-scaling experiments. \textit{Left} column represents tests on one billion-row dataset with all systems, while the \textit{Right} column represents a smaller 100 million-row dataset with \cylon{}, Dask, and Spark systems. \cylon{} was using the UCX/UCC \cite{shamis2015ucx} communicator, as it shows the best distributed performance. 

    Unfortunately, several challenges were encountered with running tests on Ray Datasets. It only supports unary operators (single input) currently. Therefore it has been omitted from \textit{Join} experiments. Moreover, Ray \texttt{groupby} did not complete within 3 hours, and \texttt{sort} did not show presentable results. Several issues came up with Modin as well. It only supports \texttt{broadcast\_join} implementation, which performs poorly on two similar-sized dataframe \textit{Join}. Only the Ray backend worked well with the data sets. Another observation was that Modin defaults to Pandas for \textit{Sort} (ie. limited distributed scalability). 
    
    The one billion-row strong scaling timings show that \cylon{} shows better scalability compared to the rest. Dask \& Spark Datasets show commendable scalability for \textit{Join} and \textit{Sort}, however the former displays very limited scalability for \textit{GroupBy}. A 100 million row test case (right column of Figure \ref{fig:sperf}) was performed to investigate Dask \& Spark further. This constitutes a communication-bound operation because the partition sizes are smaller. This reduces the computation complexity, however, these smaller partitions need to be communicated across the same number of workers. Under these circumstances, both Dask and Spark diverge significantly at higher parallelisms, indicating limitations in their communication implementations. There was a consistent anomaly in Spark timings for 8-32 parallelism. We hope to investigate this further with the help of the Spark community. 

    We also observe that the serial performance of \cylon{} outperforms the rest consistently, which could be directly related to \cylon{}'s C++ implementation and the use of Apache Arrow format. At every parallelism, \cylon{} distributed performance is $2-4\times$ higher than Dask/Spark consistently. These results confirm the efficacy of the proposed distributed execution model in this paper.

\section{Related Work}\label{sec:rel_work}

In a previous publication, we proposed a formal framework for designing and developing high-performance data engineering frameworks that include data structures, architectures, and program models \cite{kamburugamuve2021hptmt}. Kamburugamuve et al proposed a similar big data toolkit named \textit{Twister2} \cite{kamburugamuve2020twister2}, which is based on Java. There, the authors observed that using a BSP-like environment for data processing improves scalability, and they also introduced a DF-like API in Java named \textit{TSets}. However, \cylon{} being developed in C++ enables the native performance of hardware and provides a more robust integration to Python and R. 

In parallel to \cylon{}, Totoni et al also suggested a similar HP-DDF runtime named \textit{HiFrames} \cite{totoni2017hiframes}. They primarily attempt to compile native MPI code for DDF operators using \texttt{numba}. While there are several architectural similarities between \textit{HiFrames} and \cylon{}, the latter is the only open-source high-performance distributed dataframe system available at the moment. 

Dask \cite{Dask:online,rocklin2015dask} is one of the pioneering distributed dataframe implementations out there. It provides a Pandas-like API and is built on top of the Dask distributed execution environment. CuDF \cite{cudf} extends this implementation in Dask-CuDF to provide distributed dataframe capabilities in Nvidia GPUs. Modin \cite{modin:online,petersohn2020towards} is another dataframe implementation built on top of Dask and Ray. It provides an API identical to Pandas so that existing applications can be easily ported to a distributed execution. Apache Spark \cite{spark:online,zaharia2012resilient} also provides a Pandas-like DDF named \textit{Pandas on Spark}.

In addition to these systems, we would also like to recognize some exciting new projects. Velox is a C++ vectorized database acceleration library managed by the Meta Inc. incubator \cite{pedreiravelox}. Currently, it does not provide a DF abstraction, but still offers most of the operators shown in Figure \ref{fig:cylon_modin}. Photon is another C++-based vectorized query engine developed by Databricks \cite{behm2022photon} that enables native performance to the Apache Spark ecosystem. Unfortunately, it has yet to be released to the open-source community. Substrait is another interesting model that attempts to produce an independent description of data compute operations \cite{substrai7:online}.

\section{Limitations and Future Work}
\label{sec:fut}

\begin{sloppypar}
\cylon{} currently covers about 30\% of the Pandas API, and more distributed operators are being added, significantly, \textit{Window} operators. Furthermore, the cost model for evaluating dataframe operator patterns has allowed us to identify areas of improvement. For example, communication operations could be improved by introducing algorithms that have lower latency costs.
\end{sloppypar}

Additionally, in Section \ref{sec:comm-comp} we saw significant time being spent on communication. These observations can be further analyzed using MPI profiler tools (eg. TAU - Tuning and Analysis Utilities, LLNL mpiP, etc.) and distributed debugging tools (eg. Arm/Linaro DDT, etc). Some of these tools are available in the Summit supercomputer, which could give an in-depth look at the communication bottlenecks. In modern CPU hardware, we can perform computation while waiting on communication results. Since an operator consists of sub-operators arranged in a DAG, we can exploit \textit{pipeline parallelism} by overlapping communication and computation. Furthermore, we can also change the granularity of a computation such that it fits into CPU caches. We have made some preliminary investigations on these ideas, and we were able to see significant performance improvements for \cylon{}. 

Providing fault tolerance in an MPI-like environment is quite challenging, as it operates under the assumption that the communication channels are alive throughout the application. This means providing communication-level fault tolerance would be complicated. However, we are planning to add a checkpointing mechanism that would allow a much coarser-level fault tolerance. Load imbalance (especially with skewed datasets) could starve some processes and might reduce the overall throughput. To avoid such scenarios, we are working on a sample-based repartitioning mechanism.

\section{Conclusion}

We recognize that today's data science communication operations could be improved by introducing algorithms that have lower latency costs. The data science community requires scalable solutions to meet its ever-growing data demand. Dataframes are at the heart of such applications, and in this paper, we discussed a cost model for evaluating the performance of distributed dataframe operator patterns introduced in our prior publication \cite{perera2022high}. We also extended the execution model described in the previous work, by introducing a communication model. With these additions, we strongly believe we have presented a comprehensive execution model for distributed dataframe operators in distributed memory environments. Additionally, we presented \cylon{}, a reference runtime developed based on these concepts. We use the proposed model to analyze the communication and computation performance and identify bottlenecks and areas of improvement. We also showcased the importance of this work by conducting large-scale experiments on the ORNL Summit supercomputer where it showed admirable scalability in both strong and weak scaling experiments. \cylon{} also showed superior scalability compared to the state-of-the-art distributed dataframe systems, which further substantiates the effectiveness of the execution model presented in this paper.

\section*{Acknowledgments}
We gratefully acknowledge the support of NSF grants 2210266 (CINES) and 1918626 (GPCE).

\bibliographystyle{elsarticle-num}
\bibliography{ref.bib}

\begin{thebibliography}{10}
\expandafter\ifx\csname url\endcsname\relax
  \def\url#1{\texttt{#1}}\fi
\expandafter\ifx\csname urlprefix\endcsname\relax\def\urlprefix{URL }\fi
\expandafter\ifx\csname href\endcsname\relax
  \def\href#1#2{#2} \def\path#1{#1}\fi

\bibitem{perera2022high}
N.~Perera, S.~Kamburugamuve, C.~Widanage, V.~Abeykoon, A.~Uyar, K.~Shan,
  H.~Maithree, D.~Lenadora, T.~A. Kanewala, G.~Fox, High performance dataframes
  from parallel processing patterns, arXiv preprint arXiv:2209.06146.

\bibitem{hadoop:online}
Apache hadoop, \url{https://hadoop.apache.org/}.

\bibitem{dean2008mapreduce}
J.~Dean, S.~Ghemawat, Mapreduce: simplified data processing on large clusters,
  Communications of the ACM 51~(1) (2008) 107--113.

\bibitem{spark:online}
Apache spark™ - unified engine for large-scale data analytics,
  \url{https://spark.apache.org/}.

\bibitem{flink:online}
Apache flink: Stateful computations over data streams,
  \url{https://flink.apache.org/}.

\bibitem{mckinney2011pandas}
W.~McKinney, et~al., Pandas: a foundational python library for data analysis
  and statistics, Python for High Performance and Scientific Computing 14~(9)
  (2011) 1--9.

\bibitem{valiant1990bridging}
L.~G. Valiant, A bridging model for parallel computation, Communications of the
  ACM 33~(8) (1990) 103--111.

\bibitem{fox1989solving}
G.~Fox, M.~Johnson, G.~Lyzenga, S.~Otto, J.~Salmon, D.~Walker, R.~L. White,
  Solving problems on concurrent processors vol. 1: General techniques and
  regular problems, Computers in Physics 3~(1) (1989) 83--84.

\bibitem{cylon}
Cylondata, Cylon, \url{https://github.com/cylondata/cylon}.

\bibitem{gao2021scaling}
H.~Gao, N.~Sakharnykh, Scaling joins to a thousand gpus, in: 12th International
  Workshop on Accelerating Analytics and Data Management Systems Using Modern
  Processor and Storage Architectures, ADMS@ VLDB, 2021.

\bibitem{widanage2020high}
C.~Widanage, N.~Perera, V.~Abeykoon, S.~Kamburugamuve, T.~A. Kanewala,
  H.~Maithree, P.~Wickramasinghe, A.~Uyar, G.~Gunduz, G.~Fox, High performance
  data engineering everywhere, in: 2020 IEEE International Conference on Smart
  Data Services (SMDS), IEEE, 2020, pp. 122--132.

\bibitem{cudf}
rapidsai/cudf: cudf - gpu dataframe library,
  \url{https://github.com/rapidsai/cudf}.

\bibitem{moritz2018ray}
P.~Moritz, R.~Nishihara, S.~Wang, A.~Tumanov, R.~Liaw, E.~Liang, M.~Elibol,
  Z.~Yang, W.~Paul, M.~I. Jordan, et~al., Ray: A distributed framework for
  emerging $\{$AI$\}$ applications, in: 13th USENIX Symposium on Operating
  Systems Design and Implementation (OSDI 18), 2018, pp. 561--577.

\bibitem{daskshuffle:online}
Shuffling for groupby and join — dask documentation,
  \url{https://docs.dask.org/en/stable/dataframe-groupby.html}.

\bibitem{rayshuffle:online}
Performance tips and tuning — ray 2.0.0,
  \url{https://docs.ray.io/en/latest/data/performance-tips.html}.

\bibitem{shamis2015ucx}
P.~Shamis, M.~G. Venkata, M.~G. Lopez, M.~B. Baker, O.~Hernandez, Y.~Itigin,
  M.~Dubman, G.~Shainer, R.~L. Graham, L.~Liss, et~al., Ucx: an open source
  framework for hpc network apis and beyond, in: 2015 IEEE 23rd Annual
  Symposium on High-Performance Interconnects, IEEE, 2015, pp. 40--43.

\bibitem{OpenMPI:online}
Open mpi: Open source high performance computing,
  \url{https://www.open-mpi.org/}.

\bibitem{gloo:online}
facebookincubator/gloo: Collective communications library with various
  primitives for multi-machine training.,
  \url{https://github.com/facebookincubator/gloo}.

\bibitem{PMIx:online}
Pmix | process management interface - exascale copyright 2017-2020 pmix
  community, \url{https://pmix.github.io/}.

\bibitem{petersohn2020towards}
D.~Petersohn, S.~Macke, D.~Xin, W.~Ma, D.~Lee, X.~Mo, J.~E. Gonzalez, J.~M.
  Hellerstein, A.~D. Joseph, A.~Parameswaran, Towards scalable dataframe
  systems, arXiv preprint arXiv: 2001.00888.

\bibitem{hockney1994communication}
R.~W. Hockney, The communication challenge for mpp: Intel paragon and meiko
  cs-2, Parallel computing 20~(3) (1994) 389--398.

\bibitem{culler1993logp}
D.~Culler, R.~Karp, D.~Patterson, A.~Sahay, K.~E. Schauser, E.~Santos,
  R.~Subramonian, T.~Von~Eicken, Logp: Towards a realistic model of parallel
  computation, in: Proceedings of the fourth ACM SIGPLAN symposium on
  Principles and practice of parallel programming, 1993, pp. 1--12.

\bibitem{alexandrov1997loggp}
A.~Alexandrov, M.~F. Ionescu, K.~E. Schauser, C.~Scheiman, Loggp: Incorporating
  long messages into the logp model for parallel computation, Journal of
  parallel and distributed computing 44~(1) (1997) 71--79.

\bibitem{thakur2005optimization}
R.~Thakur, R.~Rabenseifner, W.~Gropp, Optimization of collective communication
  operations in mpich, The International Journal of High Performance Computing
  Applications 19~(1) (2005) 49--66.

\bibitem{traff2014implementing}
J.~L. Tr{\"a}ff, A.~Rougier, S.~Hunold, Implementing a classic: Zero-copy
  all-to-all communication with mpi datatypes, in: Proceedings of the 28th ACM
  international conference on Supercomputing, 2014, pp. 135--144.

\bibitem{bruck1997efficient}
J.~Bruck, C.-T. Ho, S.~Kipnis, E.~Upfal, D.~Weathersby, Efficient algorithms
  for all-to-all communications in multiport message passing systems, IEEE
  Transactions on parallel and distributed systems 8~(11) (1997) 1143--1156.

\bibitem{pjevsivac2007performance}
J.~Pje{\v{s}}ivac-Grbovi{\'c}, T.~Angskun, G.~Bosilca, G.~E. Fagg, E.~Gabriel,
  J.~J. Dongarra, Performance analysis of mpi collective operations, Cluster
  Computing 10~(2) (2007) 127--143.

\bibitem{shroff2000collmark}
M.~Shroff, R.~A. Van De~Geijn, Collmark: Mpi collective communication
  benchmark, in: International Conference on Supercomputing, Citeseer, 2000,
  p.~10.

\bibitem{rabenseifner2004optimization}
R.~Rabenseifner, Optimization of collective reduction operations, in:
  International Conference on Computational Science, Springer, 2004, pp. 1--9.

\bibitem{li1993versatility}
X.~Li, P.~Lu, J.~Schaeffer, J.~Shillington, P.~S. Wong, H.~Shi, On the
  versatility of parallel sorting by regular sampling, Parallel Computing
  19~(10) (1993) 1079--1103.

\bibitem{perera2020fast}
N.~Perera, V.~Abeykoon, C.~Widanage, S.~Kamburugamuve, T.~A. Kanewala,
  P.~Wickramasinghe, A.~Uyar, H.~Maithree, D.~Lenadora, G.~Fox, A fast,
  scalable, universal approach for distributed data reductions, in:
  International Workshop on Big Data Reduction, IEEE Big Data, 2020.

\bibitem{barthels2017distributed}
C.~Barthels, I.~M{\"u}ller, T.~Schneider, G.~Alonso, T.~Hoefler, Distributed
  join algorithms on thousands of cores, Proceedings of the VLDB Endowment
  10~(5) (2017) 517--528.

\bibitem{Summit89:online}
Summit user guide - olcf user documentation, \url{https://docs.olcf.ornl.gov/}.

\bibitem{Conda:online}
Conda - conda documentation, \url{https://docs.conda.io/}.

\bibitem{PyPI:online}
Pypi - the python package index, \url{https://pypi.org/}.

\bibitem{IBMSpect59:online}
IBM, Ibm spectrum mpi - overview,
  \url{https://www.ibm.com/products/spectrum-mpi}.

\bibitem{pandas:online}
pandas - python data analysis library, \url{https://pandas.pydata.org/}.

\bibitem{perera2023high}
N.~Perera, S.~Kamburugamuve, C.~Widanage, V.~Abeykoon, A.~Uyar, K.~Shan,
  H.~Maithree, D.~Lenadora, T.~A. Kanewala, G.~Fox, High performance dataframes
  from parallel processing patterns, in: Parallel Processing and Applied
  Mathematics: 14th International Conference, PPAM 2022, Gdansk, Poland,
  September 11--14, 2022, Revised Selected Papers, Part I, Springer, 2023, pp.
  291--304.

\bibitem{perera2023towards}
D.~N. Perera, Towards scalable high performance data engineering systems, Ph.D.
  thesis, Indiana University (2023).

\bibitem{perera2023supercharging}
N.~Perera, K.~Shan, S.~Kamburugamuwe, T.~A. Kanewela, C.~Widanage, A.~Sarker,
  M.~Staylor, T.~Zhong, V.~Abeykoon, G.~Fox, Supercharging distributed
  computing environments for high performance data engineering, arXiv preprint
  arXiv:2301.07896.

\bibitem{TPCHomep1:online}
Tpc-homepage, \url{https://www.tpc.org/default5.asp}.

\bibitem{Optimizi45:online}
NVIDIA, Optimizing the deep learning recommendation model on nvidia gpus,
  \url{https://developer.nvidia.com/blog/optimizing-dlrm-on-nvidia-gpus/}.

\bibitem{kamburugamuve2021hptmt}
S.~Kamburugamuve, C.~Widanage, N.~Perera, V.~Abeykoon, A.~Uyar, T.~A. Kanewala,
  G.~Von~Laszewski, G.~Fox, Hptmt: Operator-based architecture for scalable
  high-performance data-intensive frameworks, in: 2021 IEEE 14th International
  Conference on Cloud Computing (CLOUD), IEEE, 2021, pp. 228--239.

\bibitem{kamburugamuve2020twister2}
S.~Kamburugamuve, K.~Govindarajan, P.~Wickramasinghe, V.~Abeykoon, G.~Fox,
  Twister2: Design of a big data toolkit, Concurrency and Computation: Practice
  and Experience 32~(3) (2020) e5189.

\bibitem{totoni2017hiframes}
E.~Totoni, W.~U. Hassan, T.~A. Anderson, T.~Shpeisman, Hiframes: High
  performance data frames in a scripting language, arXiv preprint
  arXiv:1704.02341.

\bibitem{Dask:online}
Dask | scale the python tools you love, \url{https://www.dask.org/}.

\bibitem{rocklin2015dask}
M.~Rocklin, Dask: Parallel computation with blocked algorithms and task
  scheduling, in: Proceedings of the 14th python in science conference, Vol.
  130, Citeseer, 2015, p. 136.

\bibitem{modin:online}
Modin, Scale your pandas workflow by changing a single line of code — modin
  0.18.0 documentation, \url{https://modin.readthedocs.io/en/stable/}.

\bibitem{zaharia2012resilient}
M.~Zaharia, M.~Chowdhury, T.~Das, A.~Dave, J.~Ma, M.~McCauly, M.~J. Franklin,
  S.~Shenker, I.~Stoica, Resilient distributed datasets: A
  $\{$Fault-Tolerant$\}$ abstraction for $\{$In-Memory$\}$ cluster computing,
  in: 9th USENIX Symposium on Networked Systems Design and Implementation (NSDI
  12), 2012, pp. 15--28.

\bibitem{pedreiravelox}
P.~Pedreira, O.~Erling, M.~Basmanova, K.~Wilfong, L.~Sakka, K.~Pai, W.~He,
  B.~Chattopadhyay, Velox: Meta’s unified execution engine.

\bibitem{behm2022photon}
A.~Behm, S.~Palkar, U.~Agarwal, T.~Armstrong, D.~Cashman, A.~Dave,
  T.~Greenstein, S.~Hovsepian, R.~Johnson, A.~Sai~Krishnan, et~al., Photon: A
  fast query engine for lakehouse systems, in: Proceedings of the 2022
  International Conference on Management of Data, 2022, pp. 2326--2339.

\bibitem{substrai7:online}
substrait-io/substrait: A cross platform way to express data transformation,
  relational algebra, standardized record expression and plans.,
  \url{https://github.com/substrait-io/substrait}.

\end{thebibliography}







\end{document}